\documentclass[letterpaper,12pt]{article}
\usepackage{color,amssymb,multirow,enumerate}

\usepackage{ifpdf}
\ifpdf
	\usepackage[pdftex]{graphicx}
	\usepackage[pdftex,unicode,implicit]{hyperref}

	\hypersetup{
  	pdftitle     = {Wormholes and naked singularities in the complete Ho\v{r}ava theory},
  	pdfkeywords  = {},
  	pdfauthor    = {J. Bellor\'{\i}n, A. Restuccia and A. Sotomayor},
  	pdfcreator   = {pdf\LaTeXe\ with package \flqq hyperref\frqq},
  	pdfproducer  = {pdf\LaTeXe\ with package \flqq hyperref\frqq},
  	pdfpagemode  = UseNone,  
  	pdffitwindow = true,  
  	unicode      = true,
  	plainpages   = true,
  	colorlinks   = true,  
  	citecolor    = blue,  
  	urlcolor     = blue,
  	linkcolor    = blue
	}

	\newcommand{\grqc}[1]{
		{\href{http://www.arXiv.org/abs/gr-qc/#1}{arXiv:gr-qc/#1}}}

	\newcommand{\arXiv}[1]{
		{\href{http://www.arXiv.org/abs/#1}{arXiv:#1}}}

\else

  \usepackage[dvips]{graphicx}
  \usepackage[unicode,implicit]{hyperref}
  
  \newcommand{\grqc}[1]{{arXiv:gr-qc/#1}}
  
  \newcommand{\arXiv}[1]{{arXiv:#1}}

\fi

\makeatletter
\@addtoreset{equation}{section}
\makeatother

\begin{document}


\begin{center}

\vspace*{2cm}
{\bf \LARGE Wormholes and naked singularities in the complete Ho\v{r}ava theory} 
\vspace*{2cm}

{\sl\large Jorge Bellor\'{\i}n,}$^{a,}$\footnote{\tt jorgebellorin@usb.ve}
{\sl\large Alvaro Restuccia}$^{a,b,}$\footnote{\tt arestu@usb.ve}
{\sl\large and Adri\'an Sotomayor}$^{c,}$\footnote{\tt asotomayor@uantof.cl}
\vspace{3ex}

$^a${\it Department of Physics, Universidad Sim\'on Bol\'{\i}var, Valle de Sartenejas,\\ 
1080-A Caracas, Venezuela.} \\[1ex]
$^b${\it Department of Physics}, $^c${\it Department of Mathematics, Universidad de Antofagasta, 1240000 Antofagasta, Chile.}

\vspace*{2cm}
{\bf Abstract}
\begin{quotation}{\small
We find the static spherically symmetric solutions (with vanishing shift function) of the complete nonprojectable Ho\v{r}ava theory explicitly, writing the space-time metrics as explicit tensors in local coordinate systems. This completes previous works of other authors that have studied the same configurations. The solutions depend on the coupling constant $\alpha$ of the $(\partial_i \ln N)^2$ term. The $\lambda =1/3$ case of the theory does not possess any extra mode, hence the range of $\alpha$ is in principle not limited by the linear stability of any extra mode. We study the full range of $\alpha$, both in the positive and negative sectors. We find the same wormhole solutions and naked singularities that were found for the Einstein-aether theory in a sector of the space of $\alpha$. There also arise wormholes in other sector of $\alpha$. Our coordinate systems are valid at the throats of the wormholes. We also find the perturbative solutions for small $\alpha$. We give this version of the solutions directly on the original radial coordinate $r$, which is particularly suitable for representing the exterior region of solutions with localized sources.
}\end{quotation}

\end{center}

\thispagestyle{empty}


\newpage
\section{Introduction}
In recent years there have been advances in establishing the consistency of the complete nonprojectable Ho\v{r}ava theory \cite{Horava:2009uw,Blas:2009qj}\footnote{The original Ho\v{r}ava theory of gravitation was formulated in Ref.~\cite{Horava:2009uw}. An important class of new terms were incorporated to the potential of the nonprojectable version by Blas, Pujol\`as and Sibiryakov in Ref.~\cite{Blas:2009qj}. These terms are needed for a renormalization procedure based on the gauge symmetry; hence we use the word \emph{complete}. We use the term ``original theory'' to refer to any nonprojectable model without the terms of \cite{Blas:2009qj}, independent of whether or not they impose the detailed balance principle. Previous ideas about the foliation-preserving diffeomorphisms as a group of gauge symmetry were used by Ho\v{r}ava for a membrane action in Ref.~\cite{Horava:2008ih}.}, as well as in exploring its viability as a candidate for an ultraviolet completion of general relativity (GR) that can be consistently quantized perturbatively. Most of the early claims about the inconsistency of the theory concern a potentially dangerous extra mode. If one makes only a preliminary comparison between the number of field variables and the number of gauge symmetries, which are less in Ho\v{r}ava theory than in GR, one could conclude that in Ho\v{r}ava theory there is one degree of freedom more than in GR. Although this can be regarded as a generic result, we recently showed \cite{Bellorin:2013zbp} that \emph{this is not true for the nonprojectable theory with the special value $\lambda = 1/3$}, where $\lambda$ is the coupling constant of the tracelike kinetic term. We showed that at this point the theory has two extra second-class constraints that eliminate the extra mode from the phase space. In turn, the extra constraints protect the value $\lambda = 1/3$ against renormalization running since other values of $\lambda$ violate the constraints. Technically, what makes the value $\lambda = 1/3$ so special is that the time derivative of the spatial metric cannot be completely solved in terms of the canonical momentum from the pure Legendre transformation; instead the extra constraints arise. This result is valid even with the whole $z=3$ potential of the theory, not only for effective Lagrangians \cite{Bellorin:2013zbp}. Thus, the complete nonprojectable Ho\v{r}ava theory at $\lambda = 1/3$ has a remarkable feature: it possesses exactly the same degrees of freedom of GR. As an immediate consequence, the theory avoids the problem of strong coupling of the extra mode (low-energy divergences of self-interaction coupling constants) \cite{Charmousis:2009tc,Papazoglou:2009fj,Blas:2009ck,Kimpton:2010xi}, since there is no extra mode. Having the same degrees of freedom of GR at all scales, the $\lambda =1/3$ theory does not experience the obstructions other theories of gravitation frequently face in recovering GR: discontinuities or ghosts. Moreover, the linear-order perturbative version of the effective theory for large distances is physically equivalent to the linearized version of GR. This implies that the effective theory propagates gravitational waves exactly in the same way as linearized GR does. Measurements that could signal gravitational waves have been recently obtained \cite{Ade:2014xna}.

In the linear-order perturbative analysis of Ref.~\cite{Park:2009hg}, the absence of the extra mode for a particular model of the original nonprojectable Ho\v{r}ava theory at $\lambda = 1/3$ with a Cotton-square term (without the $a_i = \partial_i \ln N$ terms) was found. Since the model considered there acquires conformal symmetry at the UV, one could think that the absence of the extra mode is a consequence of the approximate conformal symmetry, but the analysis of \cite{Bellorin:2013zbp} shows that this result is actually a consequence of the $\lambda = 1/3$ value. Extra gauge symmetries would lower furthermore the number of degrees of freedom in the $\lambda =1/3$ case.

Even in the theory with $\lambda \neq 1/3$, where the extra mode arises, the disastrous strong coupling could be circumvented, since it arises only  if one forces the limit $\lambda \rightarrow 1$ in order to recover GR at low energies \cite{Charmousis:2009tc}. In Ref.~\cite{Bellorin:2010je} (see also \cite{Das:2011tx}) two of us showed a way to recover GR for any $\lambda$: the second-order model having only the spatial scalar curvature in the potential is physically equivalent to GR with no need of assuming additional restrictions on the kinetic terms. Thus, the way to recover GR is to neglect all the terms in the potential except the $^{(3)}R$ term, whereas $\lambda$ is left arbitrary. This mechanism avoids the limit $\lambda \rightarrow 1$ for recovering GR.

For both cases, $\lambda = 1/3$ and $\lambda \neq 1/3$, it has been shown the consistency of the Hamiltonian formulation of the complete nonprojectable theory \cite{Bellorin:2013zbp,Donnelly:2011df,Bellorin:2011ff,Bellorin:2012di} (see also Ref.~\cite{Kluson:2010nf}). It has been fully clarified the set of constraints of the theory, together with their first- or second-class character. There is no evidence leading us to conclude that any of those constraints is an inconsistent equation for the metric variables or their conjugated momenta.\footnote{The equation used in Ref.~\cite{Henneaux:2009zb} to argue that the original nonprojectable Ho\v{r}ava theory necessarily has vanishing lapse function is actually an equation for the trace of the canonical momentum with the physically consistent solution $\pi = 0$ \cite{Bellorin:2010je}.} It has been shown that the algebra of constraints closes. All the constraints are preserved in time once consistent equations for the Lagrange multipliers associated to second-class constraints are imposed\footnote{When the original (incomplete) theory is considered with terms of higher order the algebra of constraints also closes, but there arise nonelliptic differential equations \cite{Bellorin:2010te}. In Ref.~\cite{Li:2009bg} the second-class constraints of the original theory were misinterpreted as first-class ones. That prevents to obtain the closure of their algebra.}.

These results, in particular the outstanding properties of the $\lambda = 1/3$ theory, encourage us to deepen on the physical content of the complete nonprojectable Ho\v{r}ava theory. Because of their astrophysical relevance, we devote this paper to study the (vacuum) static spherically symmetric solutions of the complete nonprojectable Ho\v{r}ava theory. Since the interest is in the large-distance physics, we use the lowest-order effective action (second order in derivatives). Unlike the symmetries of GR, the foliation-preserving diffeomorphism symmetry is not enough to eliminate the mixed time-space components of the metric once it is written in spherical coordinates (Schwarzschild coordinates). Thus, those components (the shift function) are left active if only symmetry arguments are used, representing a hair for this class of solutions. For the sake of simplicity, in this paper we only consider static spherically symmetric configurations with vanishing shift function\footnote{Static spherically symmetric solutions with active shift function were found for the original Ho\v{r}ava theory in Ref.~\cite{Capasso:2009ks}.}. The analysis covers simultaneously the $\lambda = 1/3$ and $\lambda \neq 1/3$ cases, since $\lambda$ is a multiplier of a kinetic term. Once the conditions of staticity and vanishing of the shift function are imposed, all kinetic terms vanish, such that $\lambda$ disappears from the field equations. Thus, under these conditions the field equations are the same for $\lambda = 1/3$ and $\lambda \neq 1/3$. The only coupling constant the field equations depend on is the one of the $(\partial_i \ln N )^2$ term, which is denoted by $\alpha$.

Some years before the formulation of the Ho\v{r}ava theory, Eling and Jacobson \cite{Eling:2006df} (see also \cite{Eling:2003rd}) studied the static spherically symmetric solutions of the Einstein-aether theory (EA theory), which is also a theory with preferred frame \cite{Jacobson:2000xp}. It turns out that such solutions are also the static spherically symmetric solutions of the effective action of the complete nonprojectable Ho\v{r}ava theory, since the EA theory becomes equivalent to the large-distance limit of the Ho\v{r}ava theory when the condition of hypersurface orthogonality is imposed on the unit vector of the EA theory \cite{Blas:2009ck,Jacobson:2010mx,Jacobson:2013xta}. This condition is met for the static spherically symmetric configurations \cite{Eling:2006df}.
Eling and Jacobson \cite{Eling:2006df} studied the solutions in a restricted range for the coupling constant equivalent to $\alpha$. In that range they could find the solutions analytically in terms of a function of the radius. They used such a function as a parameter over which the solution can be explicitly expressed. They found that the solutions have a wormholelike geometry, with two spatial branches joined by a throat (an Einstein-Rosen bridge). These solutions are not black holes.

Directly on the side of the complete nonprojectable Ho\v{r}ava theory, an analysis of the static spherically solutions was done by Kiritsis in Ref.~\cite{Kiritsis:2009vz} (in \cite{Kiritsis:2009rx} there is a related study on the original theory). Motivated by the linear stability of the (possible) extra mode \cite{Blas:2009qj}, he concentrated the analysis on the $0 < \alpha < 2$ range. This is the same range studied for the solutions of the EA theory in Ref.~\cite{Eling:2006df}. Kiritsis also analyzed the solutions in the $\alpha = 2$ and $\alpha > 2$ ranges. He got the solutions in a similar fashion to \cite{Eling:2006df}: expressing them in terms of a function of the radius and then using the function as the parameter that controls the configurations. In addition, there is vast literature about solutions of the original Ho\v{r}ava theory, both in the projectable and nonprojectable versions. Most of these studies are perturbative approaches. Exact computations on finding static spherically symmetric solutions can be found in Refs.~\cite{Lu:2009em,Kehagias:2009is}.

In this paper we shall make further developments on analyzing the field equations for static spherically symmetric solutions. Our aim is to find the solutions in a closed way, such that the space-time metric can be explicitly written as a tensor in concrete local coordinate systems. Furthermore, since the $\lambda = 1/3$ theory does not possess any extra mode, the range of $\alpha$ is not restricted \emph{a priori} by physical features associated to the extra mode if one is interested in the dynamics of the $\lambda = 1/3$ theory. Therefore, we extend the analysis of the static spherically solutions to the full range of $\alpha$, including negative values. In Section \ref{sec:exactsolutions} we shall show that the field equations can be exactly and explicitly solved in a closed way. We shall perform a managing of the field equations aimed to recover explicitly the constraints of the theory. The central step of the procedure will consist of extracting a purely \emph{algebraic} field equation. This equation can be easily solved according to two different cases of $\alpha$; the solutions arising as one-parameter families on each case. We shall use the corresponding parameter as a transformed radial coordinate, such that the final solution will be explicitly written in terms of the transformed radius. The final exact and explicit expressions will help us to furthermore understand the properties of the solutions.

In order to further explore the solutions, in Section \ref{sec:perturbations} we present a different and convenient approach for solving the equations: the perturbative approach. We shall show how the field equations can be solved approximately by assuming an small $\alpha$. Specifically, the analysis will be done at linear order in $\alpha$. The explicit expression we shall find for the perturbative solution has the advantage of being given directly in the original radius $r$ of the spherical coordinates. We shall discuss the spatial ranges of validity of the perturbative solutions. Section 5 contains further discussion and conclusions.


\section{The conditions of staticity and spherical symmetry} 
The action of the complete, nonprojectable Ho\v{r}ava theory is written in terms of the Arnowitt-Deser-Misner (ADM) variables $g_{ij}$, $N$ and $N_i$ as
\begin{equation}
 S = \int dt d^3x \sqrt{g} N 
       ( G^{ijkl} K_{ij} K_{kl} - \mathcal{V} ),
\label{lagrangianaction}
\end{equation}
where
\begin{eqnarray}
K_{ij} & = & \frac{1}{2N} ( \dot{g}_{ij} - 2 \nabla_{(i} N_{j)} ) \,,
\\[1ex]
G^{ijkl} & = &
\frac{1}{2} ( g^{ik} g^{jl} + g^{il} g^{jk} ) 
- \lambda g^{ij} g^{kl} \,,
\end{eqnarray}
and $\mathcal{V}$ is the potential, which depends explicitly on the curvature tensors, the vector $a_i \equiv \partial_i \ln N$ and derivatives of them. In the nonprojectable theory the lapse function $N$ is regarded as a function of both time and space. We specialize our analysis to the large-distance effective action, which is formed with the quadratic potential
\begin{equation}
 \mathcal{V}^{(2)} = - R - \alpha a_i a^i \,,
\label{quadraticpotential}
\end{equation}
where $\alpha$ is a coupling constant. One may put an additional, undetermined coupling constant for the $R$ term since it is covariant by itself under the gauge symmetry of the theory, which is the foliation-preserving diffeomorphisms group, with no need of mixing with the kinetic term. Since such a constant can be always absorbed in the free theory by rescaling the time, we omit it in our discussion.\footnote{If this free constant is kept, the only effect it would have on static solutions is to replace the coupling constant $\alpha$ by the ratio of $\alpha$ and it.}

The corresponding equations of motion, which are derived by taking variations with respect to $g_{ij}$, $N$ and $N_i$, are given, respectively, by
\begin{eqnarray}
 \frac{1}{\sqrt{g}} \frac{\partial}{\partial t} 
                   ( \sqrt{g} G^{ijkl} K_{kl} )
 + 2 N ( K^{ik} K_k{}^j - \lambda K K^{ij} )
 - \frac{1}{2} N g^{ij} G^{klmn} K_{kl} K_{mn} 
& & \nonumber \\
 + 2 \nabla_k ( G^{kmn(i} K_{mn} N^{j)} )
 - \nabla_k ( G^{mnij} K_{mn} N^k )
 + N (R^{ij} - \frac{1}{2} g^{ij} R )
& & \nonumber \\ 
 - ( \nabla^i \nabla^j N - g^{ij} \nabla^2 N )
 + \alpha N^{-1} ( \nabla^i N \nabla^j N 
    - \frac{1}{2} g^{ij} \nabla_k N \nabla^k N )
& = & 0 \,,
\label{einstein}
\\
 G^{ijkl} K_{ij} K_{kl} - R 
 + 2 \alpha N^{-2} ( N \nabla^2 N - \frac{1}{2} \nabla_i N \nabla^i N ) 
& = & 0 \,,
\label{hamiltonianconstrainlagrange}
\\
\nabla_i ( G^{ijkl} K_{kl} ) &=& 0 \,.
\label{momentunconstrainlagrange}
\end{eqnarray}
 
We proceed to evaluate systematically the field equations for static and spherically symmetric configurations, starting with the condition of staticity. There is an important difference in the role the shift function $N_i$ has for static spherically symmetric configurations of Ho\v{r}ava theory with respect to GR. As it is well known, in GR the only nonzero component of $N_i$ for static spherically symmetric metrics, written in spherical coordinates, can be absorbed by redefining the time in a spatial-dependent way. However, in Ho\v{r}ava theory such a coordinate transformation is not allowed. Once the general static spherically symmetric metrics are written in spherical coordinates, unavoidably the radial component of the shift function remains as an arbitrary function \cite{Kiritsis:2009rx}. In spite of this, we remark  that the shift function $N_i$ \emph{is not a true functional degree of freedom} of Ho\v{r}ava theory. $N_i$ is the Lagrange multiplier of the momentum constraint which generates the spatial diffeomorphisms. It can always be set equal to zero by choosing an appropriated spatial coordinate system. The obstruction in being absorbed in static spherically symmetric metrics is a mere consequence of choosing spherical coordinates. One may write general static spherically symmetric metrics in another coordinate system without shift function but with an arbitrary function in the spatial sector. That is, one may translate the freedom parametrized in $N_i$ to the pure spatial sector. 

For the sake of simplicity, in this analysis we shall consider only static configurations with $N_i=0$ (in spherical coordinates). This implies $K_{ij} =0$. Under these considerations Eq.~(\ref{momentunconstrainlagrange}) is automatically solved, whereas Eqs.~(\ref{einstein}) and (\ref{hamiltonianconstrainlagrange}) reduce to
\begin{eqnarray}
 N (R^{ij} - \frac{1}{2} g^{ij} R )
 - ( \nabla^i \nabla^j N - g^{ij} \nabla^2 N )
 + \alpha N^{-1} ( \nabla^i N \nabla^j N 
    - \frac{1}{2} g^{ij} \nabla_k N \nabla^k N )
& = & 0 \,, \hspace*{3em}
\label{eisnteinstatic}
\\
 R - 2 \alpha N^{-2} ( N \nabla^2 N 
         - \frac{1}{2} \nabla_i N \nabla^i N ) 
& = & 0 \,.
\label{hamiltoniancstatic}
\end{eqnarray}
Equation (\ref{hamiltoniancstatic}) and the trace of (\ref{eisnteinstatic}) are equivalent to the system
\begin{eqnarray}
  R - \left( \alpha + 2 \right) N^{-1} \nabla^2 N 
  + \alpha N^{-2} \nabla_i N \nabla^i N & = & 0  \,,
\label{eqtraceprev1}
\\
 \left( \alpha - 2 \right) \nabla^2 N  & = & 0 \,.
\label{eqtraceprev2}
\end{eqnarray}

Let us concentrate first on the case $\alpha \neq 2$, leaving the study of the special case $\alpha = 2$ to the end of Section \ref{sec:exactsolutions}. If $\alpha \neq 2$, Eqs.~(\ref{eqtraceprev1}) and (\ref{eqtraceprev2}) are equivalent to
\begin{eqnarray}
  R + \alpha N^{-2} \nabla_i N \nabla^i N &=& 0 \,,
\label{hamiltonianfin}
\\
 \nabla^2 N &=& 0 \,.
\label{nablaN}
\end{eqnarray}
The equation of motion (\ref{eisnteinstatic}), after including (\ref{hamiltonianfin}) and (\ref{nablaN}) in it, yields
\begin{equation}
 R^{ij} - N^{-1} \nabla^i \nabla^j N 
   + \alpha N^{-2} \nabla^i N \nabla^j N = 0 \,.
\label{eomstaticlag}
\end{equation}
The system of equations (\ref{hamiltonianfin} - \ref{eomstaticlag}) is equivalent to the field equations (\ref{eisnteinstatic}) and (\ref{hamiltoniancstatic}) (one may derive one of the Eqs.~(\ref{hamiltonianfin}) or (\ref{nablaN}) by combining the other with (\ref{eomstaticlag})). Under the Hamiltonian formalism, Eqs.~(\ref{hamiltonianfin}) and (\ref{nablaN}) are deduced directly from constraints in the theory with $\lambda = 1/3$. In the case $\lambda \neq 1/3$, these equations are deduced from combining one constraint with the equations of motion. In Appendix A we show how this works in the Hamiltonian formalism.

Next we evaluate the equations of motion (\ref{hamiltonianfin} - \ref{eomstaticlag}) for static, spherically symmetric configurations with vanishing shift function given by the following ansatz in spherical coordinates:
\begin{equation}
 N = N(r) 
\,,\hspace{2em} 
 ds_{(3)}^2 = {\displaystyle\frac{dr^2}{f(r)} + r^2 d\Omega_{(2)}^2} \,.
\label{ansatz}
\end{equation}
Equations (\ref{hamiltonianfin}) and (\ref{nablaN}) yield, respectively,
\begin{eqnarray}
 r f' + f - 1  
 - \frac{\alpha}{2} r^2 f \left( \frac{ N'}{N} \right)^2 
 &=& 0 \,,
\label{css2}
\\
( r^2 \sqrt{f} N' )' &=& 0 \,.
\label{css1}
\end{eqnarray}
All off-diagonal components of the equation of motion (\ref{eomstaticlag}) vanish. The $rr$ and $\theta\theta$ components become, respectively,
\begin{eqnarray}
 \frac{f'}{r f} + \frac{N''}{N} + \frac{f' N'}{2 f N} 
   - \alpha \left( \frac{N'}{N} \right)^2 &=& 0 \,,
\label{eom1}
\\
 \frac{1}{2} r f' + f - 1 + \frac{ r f N'}{N} &=& 0 \,,
\label{eom2}
\end{eqnarray}
and the $\phi\phi$ component is equivalent to the $\theta\theta$ component. We notice that Eq.~(\ref{eom1}) is a linear combination of Eqs.~(\ref{css2}), (\ref{css1}) and (\ref{eom2}). The field equations reduce then exactly to (\ref{css2}), (\ref{css1}) and (\ref{eom2}). The counting of the independent initial data yields that the general solution possesses two arbitrary integration constants. Indeed, the system (\ref{css2} - \ref{eom2}) is manifestly invariant under scalings of $N$ and $r$. As in GR, the integration constant associated to scalings of $N$ can be absorbed by scaling the time coordinate, which is equivalent to impose the boundary condition $N|_{r=\infty} = 1$. Therefore, the general solution has one physical integration constant.

We close this section by commenting that the field equations (\ref{css2} - \ref{eom2}) are also valid for a theory that in addition has a term proportional to the square of the Cotton tensor, which was the original $z=3$ term Ho\v{r}ava considered \cite{Horava:2009uw}. Since three-dimensional spherically symmetric metrics are conformally flat, their Cotton tensor vanishes; hence, in the field equations all terms coming from the $(\mbox{Cotton})^2$ term trivially vanish. This fact gives further relevance to the spherically symmetric solutions in Ho\v{r}ava theory, since they arise in a more complete model in  the sense that it has both the lower-order terms dominant at large distances and a high-order term that improves the renormalizability at microscopic scales.

\section{Exact solutions}
\label{sec:exactsolutions}
In this section we focus ourselves on the exact solutions of the field equations (\ref{css2}), (\ref{css1}) and (\ref{eom2}). This will be achieved by appealing to suitable local coordinate systems on each of the cases for $\alpha$ we shall meet.

From Eq.~(\ref{css1}) we have
\begin{equation}
 r^2 \sqrt{f} N' = C \,,
\label{rfnprima}
\end{equation}
where $C$ is an integration constant. By a linear combination of Eqs.~(\ref{css2}), (\ref{eom2}) and (\ref{rfnprima}) we obtain the equation
\begin{equation}
 f - 1 + \frac{2 C \sqrt{f}}{r N} 
 + \frac{\alpha}{2} \left( \frac{C}{r N} \right)^2 = 0 \,.
\label{algebraiceq}
\end{equation}
Notice that this is an algebraic equation and it is quadratic in $\sqrt{f}$ and $(r N)^{-1}$. This is the key algebraic equation we mentioned in the Introduction. If we use the notation 
\begin{equation}
\beta \equiv \sqrt{\left|1 - \frac{\alpha}{2}\right|} \,, 
\end{equation}
this equation can be rewritten as one of the following two possibilities according to the value of $\alpha$:
\begin{eqnarray}
 \left( \sqrt{f} + \frac{C}{r N} \right)^2 
 - \left( \frac{\beta C}{r N} \right)^2 &=& 1 
\hspace{2em} \mbox{if} \hspace{2em} \alpha < 2 \,,
\label{bnegative}
\\
\left( \sqrt{f} + \frac{C}{r N} \right)^2 
 + \left( \frac{\beta C}{r N} \right)^2 &=& 1 
\hspace{2em} \mbox{if} \hspace{2em} \alpha > 2 \,.
\label{bpositive}
\end{eqnarray}
In the following subsections we analyze these two possibilities separately and at the end we complete with the $\alpha = 2$ case. We remark that after solving Eqs.~(\ref{rfnprima}) and (\ref{algebraiceq}) it is straightforward to check that the remaining field equation is identically satisfied.

\subsection{Case $\alpha < 2$}
We can give the most general solution of Eq.~(\ref{bnegative}) in terms of one-parameter families of solutions,
\begin{equation}
 \frac{\beta C}{r N} = s_1 \sinh{\chi} \,,
\hspace{2em}
 \sqrt{f} + \frac{C}{r N} = s_2 \cosh{\chi} \,,
\label{parametricsolution}
\end{equation}
where $\chi$ is an arbitrary parameter and $s_1$, $s_2$ are evaluated on any combination of signs $\pm 1$. Now, our aim is to regard the variable $\chi$ as a new radial coordinate such that the solutions for $N$ and $f$ can be explicitly expressed as functions of $\chi$. $\chi$, as well as $s_1$, $s_2$ and $C$ are restricted by positiveness of the field variables. We first note that, in order to preserve positiveness of $r$ and $N$, we must avoid changes of sign in $\sinh{\chi}$. We take the sector $\chi\in (0,+\infty)$; in Appendix \ref{apendix:equivalences} we show that the final solutions in the sector $\chi\in (-\infty,0)$ are equivalent to the ones in $\chi\in (0,+\infty)$. Next, we have the condition $\sqrt{f} > 0$, which translates itself into
\begin{equation}
 - s_1 \beta^{-1} \sinh{\chi} + s_2 \cosh{\chi} > 0 \,.
\end{equation}
There are four possibilities in the range $\chi\in (0,+\infty)$, each one valid for a specific sign of $C$ according to (\ref{parametricsolution}),
\begin{enumerate}[i]
 \item $s_1 = +1$, $s_2 = +1$ \,:\hspace{1em} is a solution in the range $0 < \tanh{\chi} < \beta$ with $C > 0$.

 \item $s_1=+1$, $s_2=-1$ \,:\hspace{1em} is not a solution.

 \item $s_1=-1$, $s_2=+1$ \,:\hspace{1em} is a solution for all $\chi \in (0,+\infty)$ with $C<0$.

 \item $s_1=-1$, $s_2=-1$ \,:\hspace{1em} is a solution in the range $\beta < \tanh{\chi} < \infty$ with $C < 0$.
\end{enumerate}

Since we are in the $\alpha < 2$ range, we have that the GR limit $\alpha = 0$ corresponds to $\beta = 1$. $\beta < 1$ corresponds to $0 < \alpha < 2$ and $\beta > 1$ to $\alpha < 0$. This implies that for $\alpha \leq 0$ the domain of validity of solution (iv) is an empty set whereas solution (i) is valid in the full range $\chi\in (0,+\infty)$. For $0 < \alpha < 2$ the domains of solutions (i) and (iv) are both nonempty sets. They do not intersect themselves and their union together with the point $\hat{\chi}$ given by
\begin{equation}
 \tanh{\hat{\chi}} = \beta
\label{throat}
\end{equation}
constitutes the full range $\chi\in (0,+\infty)$. We then see that there arises a further refinement of the range of $\alpha$: the solutions behave in different ways among the ranges $\alpha < 0$, $\alpha = 0$ and $0 < \alpha < 2$. We stress that these ranges have not arisen as consequence of the physics of any extra mode. They are a feature of the field equations for static spherically symmetric configurations.

Thus, the above list leaves us with two solutions in the $\alpha = 0$ and $\alpha < 0$ cases and three solutions in the $0 < \alpha < 2$ case. For the $0 < \alpha < 2$ range, however, the domains of validity of solutions (i) and (iv) suggest that they can be joined to form a single solution. This is just what happens, as we are going to see shortly. Notice that solutions (i) and (iv) have the same relative sign, $s \equiv s_1 s_2 = +1$, whereas solution (iii) has $s = -1$. We are going to see that the final solutions only depend on the relative sign $s$ and have a global (i.e., everywhere valid), positive, physical integration constant once they are written in the $\chi$ coordinate.

We may obtain a relation between $r$ and $\chi$ valid for any $s_1$, $s_2$ and $C$ in the full range $\alpha <2$. By combining Eqs.~(\ref{rfnprima}) and (\ref{parametricsolution}) we get the equation
\begin{equation}
 \left( \frac{\cosh\chi}{\sinh\chi} - \frac{s}{\beta} \right) 
     \frac{d\chi}{d r} 
   = - \frac{1}{r} \,.
   \label{diffeqcoord}
\end{equation}
This equation can be integrated straightforwardly, yielding
\begin{equation}
 r = \frac{k e^{s \chi / \beta}}{\sinh{\chi}} \,,
\label{coordinatetransf}
\end{equation}
where $k$ is an integration constant subject to $k > 0$ by consistency. Relation (\ref{coordinatetransf}) allows us to regard $\chi$ as a new radial coordinate. For the two cases of $s$ the value $\chi = 0$ corresponds to the spatial infinity $r = \infty$. 

By putting the transformation (\ref{coordinatetransf}) back into Eqs.~(\ref{parametricsolution}) we obtain $N$ and $\sqrt{f}$ as explicit functions of $\chi$, 
\begin{equation}
 N = s_1 \frac{ \beta C }{ k } e^{ - s \chi / \beta} \,,
 \hspace{2em}
 \sqrt{f} = s_1 ( s \cosh\chi - \beta^{-1} \sinh\chi ) \,.
\label{presolutions}
\end{equation}
These expressions give the three solutions listed above explicitly in terms of the coordinate $\chi$. Note that for any of the three solutions $s_1 C$ is nothing but $|C|$, since in any case the sign of $C$ must be compensated with $s_1$.

Now, we proceed to show that in the $0 < \alpha < 2$ range solutions (i) and (iv), which have $s = +1$, can be smoothly joined. The joining point is $\hat{\chi}$ (\ref{throat}) and at $\hat{\chi}$ the function $f$ given in (\ref{presolutions}) is continuous and equal to zero. This is just a coordinate singularity associated to the fact that $f$ is a metric component in the coordinate $r$, as we are going to see. Next, if we adjust both the absolute value of $C$ and the value of $k$ for solutions (i) and (iv), $N$ is also continuous at $\hat{\chi}$ (but not vanishing).

Therefore, there are two solutions of the field equations in the whole $\alpha < 2$ case: the $s=+1$ solution, which is either the joining of solutions (i) and (iv) or just solution (i) if $0 < \alpha < 2$ or $\alpha \leq 0$ respectively, and the $s=-1$ solution, which is solution (iii) for all $\alpha < 2$. Actually, for both solutions functions $N$ and $f$ are given in global expressions in the whole range $\chi \in (0,+\infty)$. Indeed, $|C|$ and $k$ are the unique, global, integration constants the solutions have when they are written in the $\chi$ coordinate. As we discussed above, only one integration constant has physical meaning. The role of the sign $s_1$ in front of the the ``unjoined'' expression of $\sqrt{f}$ in (\ref{presolutions}) for the $s=+1$ solution is equivalent to set $\sqrt{f}$ equal to the absolute value of the combination $\cosh\chi - \beta^{-1} \sinh\chi$ for all $\chi \in (0,+\infty)$. The everywhere valid expressions of the two solutions in $\chi\in (0,+\infty)$ are
\begin{equation}
 N = \frac{ \beta C }{ k } e^{ - s \chi / \beta} \,,
 \hspace{2em}
 f = ( \beta^{-1} \sinh\chi - s \cosh\chi )^2\,,
 \label{nffinalalphal2}
\end{equation}
where $C > 0$. Here and in the following we write the global integration constant $|C|$ just as a positive $C$ since we keep the solution in the $\chi$ coordinate, but the reader should keep in mind that this $C$ is different to the local integration constants arising in previous equations.\footnote{The constant $C$ arising from (\ref{rfnprima}) to (\ref{presolutions}) can be regarded as the local integration constant needed to cast the solutions in terms of the coordinate $r$. Necessarily it changes its sign among some sectors of the solutions.} Thus, the two solutions have a global physical integration constant and each solution is determined by the choice of $s = \pm 1$. Since expressions in (\ref{nffinalalphal2}) are $C^{\infty}$ in $\chi\in (0,+\infty)$, the union of solutions (i) and (iv) for $0 < \alpha < 2$ is completely smooth. We impose the boundary condition  $N^2|_{r=\infty} = 1$, which fixes the integration constant $k$ to $k = \beta C$. 

By combining the functions (\ref{nffinalalphal2}) with relations (\ref{diffeqcoord}) and  (\ref{coordinatetransf}) we may write the space-time metric as an explicit tensor in terms of the radial coordinate $\chi$, obtaining the two metrics,
\begin{equation}
 ds_{(4)}^2 = 
 - e^{- 2 s \chi / \beta} dt^2
 + \frac{ (\beta C)^2 e^{2 s \chi / \beta}}{\sinh^4{\chi}}
    ( d\chi^2 + \sinh^2{\chi} d\Omega_{(2)}^2 ) \,,
\label{metricalphalower2}
\end{equation}
each one determined by the choice of $s$. Both solutions are valid in the range $\chi \in (0,+\infty)$. 

Once the spatial part of the metric is written in the $\chi$ coordinate, it arises explicitly as a conformal equivalent of the metric of the 3-hyperboloid in hyperbolic coordinates, $ds^2_{H^3} = d\chi^2 + \sinh^2{\chi} d\Omega_{(2)}^2$. This is the metric induced on the hyperboloid when immersed into a flat Lorentzian $\mathbb{R}^4$ ambient. Since both the hyperboloid and the (spatial) solutions are conformally flat, there always exist local coordinate systems under which one metric can be explicitly and locally written as a conformally transformed of the other one. With relation (\ref{coordinatetransf}) we have found a coordinate system that realizes this conformal equivalence explicitly. The conformal factor, however, changes greatly the geometry of the spatial part of the solutions with respect to the 3-hyperboloid, as we are going to see in the following. 

From the explicit expression (\ref{metricalphalower2}) of the solutions in the coordinate $\chi$ we can compute curvature tensors. The nonzero components of the four-dimensional Riemann tensor are
\begin{equation}
\begin{array}{l}
^{(4)}\!R_{t \chi \chi}{}^{t} =
 - 2 (\beta \sinh{\chi})^{-1} (\beta^{-1} \sinh{\chi} - s \cosh{\chi}) \,,
\\[2ex]
^{(4)}\!R_{t \theta \theta}{}^{t} = 
 - {\displaystyle\frac{1}{2}} \sinh^2{\chi} R_{t \chi \chi}{}^{t} \,,
\hspace{2em}
^{(4)}\!R_{t \phi \phi}{}^{t} = \sin^2{\theta}\: {}^{(4)}\!R_{t \theta \theta}{}^{t} \,, 
\\[2ex]
^{(4)}\!R_{\chi\theta\theta}{}^{\chi} =
\beta^{-1} \sinh{\chi} (\beta \sinh{\chi} - s\cosh{\chi}) \,,
\hspace{2em}
^{(4)}\!R_{\chi \phi \phi}{}^{\chi} = 
 \sin^2{\theta}\:{}^{(4)}\!R_{\chi\theta\theta}{}^{\chi}  \,, \\[2ex]
^{(4)}\!R_{\theta \phi \theta}{}^{\phi} =
- \beta^{-2}( 1 - (\beta \sinh{\chi} - s\cosh{\chi})^2 ) \,.
\end{array}
\end{equation}
The 4D and 3D Ricci scalars are equal to
\begin{equation}
 ^{(4)}\!R = R =
 - \frac{\alpha}{\beta^4 C^2} e^{-2 s \chi / \beta} \sinh^4{\chi} \,.
\label{ricciscalar1}
\end{equation}

Let us start the study of the geometry of solutions (\ref{metricalphalower2}) by taking the asymptotic limit, which can be done simultaneously for both $s$. The radial coordinate $\chi$ is bad frame to study asymptotic behavior, since the metric components (\ref{metricalphalower2}) diverge at $\chi = 0$. However, we can observe the solutions near the spatial infinity by coming back to the original radial coordinate in an approximate way since the coordinate transformation (\ref{coordinatetransf}) can be inverted in the limit when $\chi \rightarrow 0$ and simultaneously $r \rightarrow \infty$. The linearized version of the transformation (\ref{coordinatetransf}) is
\begin{equation}
 \frac{1}{r} = \frac{\chi}{\beta C} \,,
\end{equation}
which is valid for both $s$. Using this transformation in (\ref{nffinalalphal2}) we obtain the expanded version of $N^2$ and $f$ at linear order in $r^{-1}$,
\begin{equation}
 N^2 = f = 1 - s \frac{2C}{r} \,.
\label{expansions}
\end{equation}
Thus, we have that both solutions are asymptotically flat for any $\alpha$ in the range $\alpha < 2$. The sign of the external mass is given by the choice of $s$. In Section \ref{sec:perturbations} we shall present the asymptotic expansion of the exact solutions up to $1/r^3$ order. As a consequence of the asymptotic flatness, the Ricci scalars given in (\ref{ricciscalar1}) vanish at $\chi = 0$ for both cases of $s$. Note that this behavior departs greatly from the 3-hyperboloid, which is a manifold of constant, negative, curvature.

\begin{figure}[t]
 \begin{center}
 \includegraphics[scale=0.40]{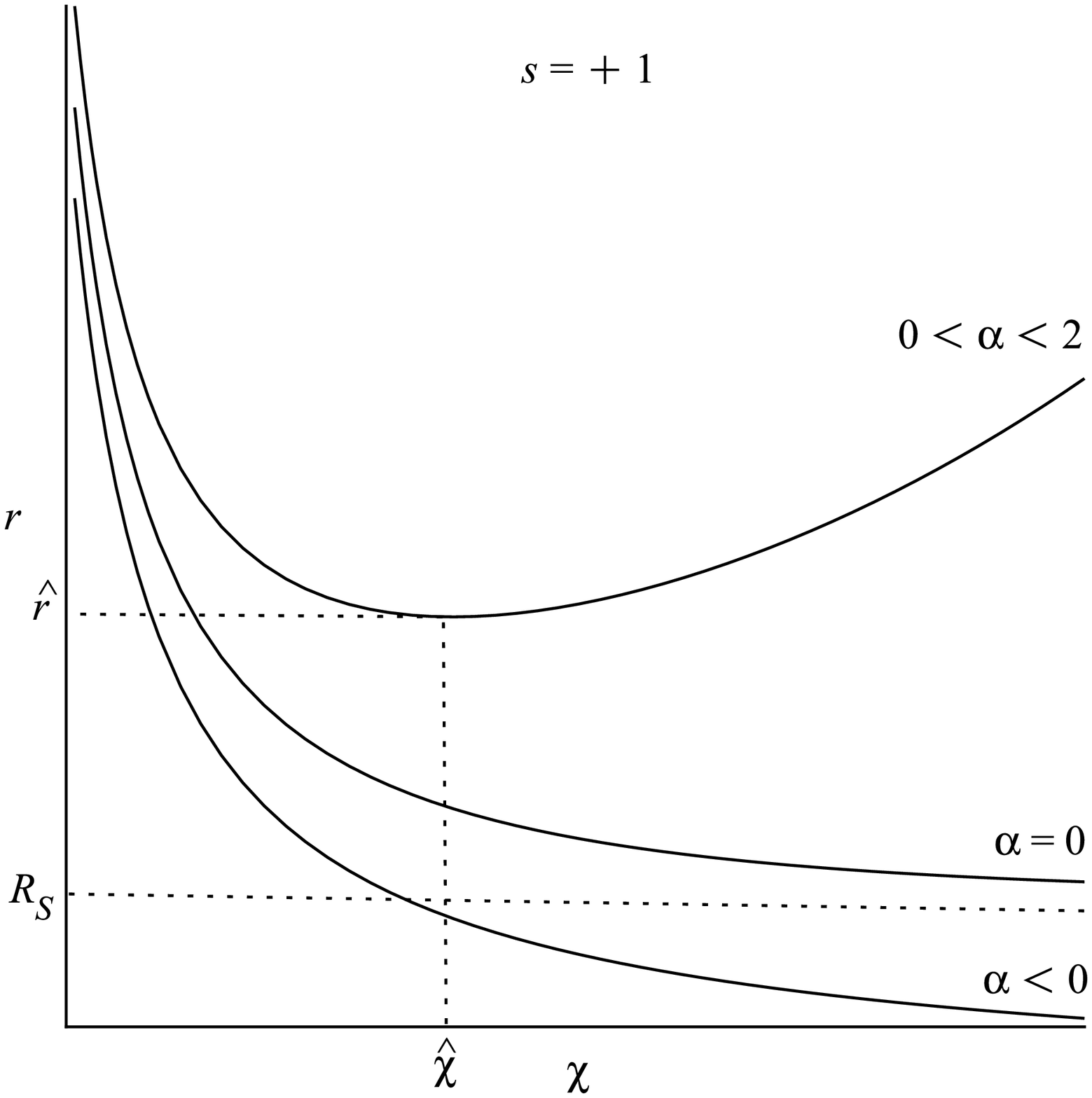}
 \includegraphics[scale=0.40]{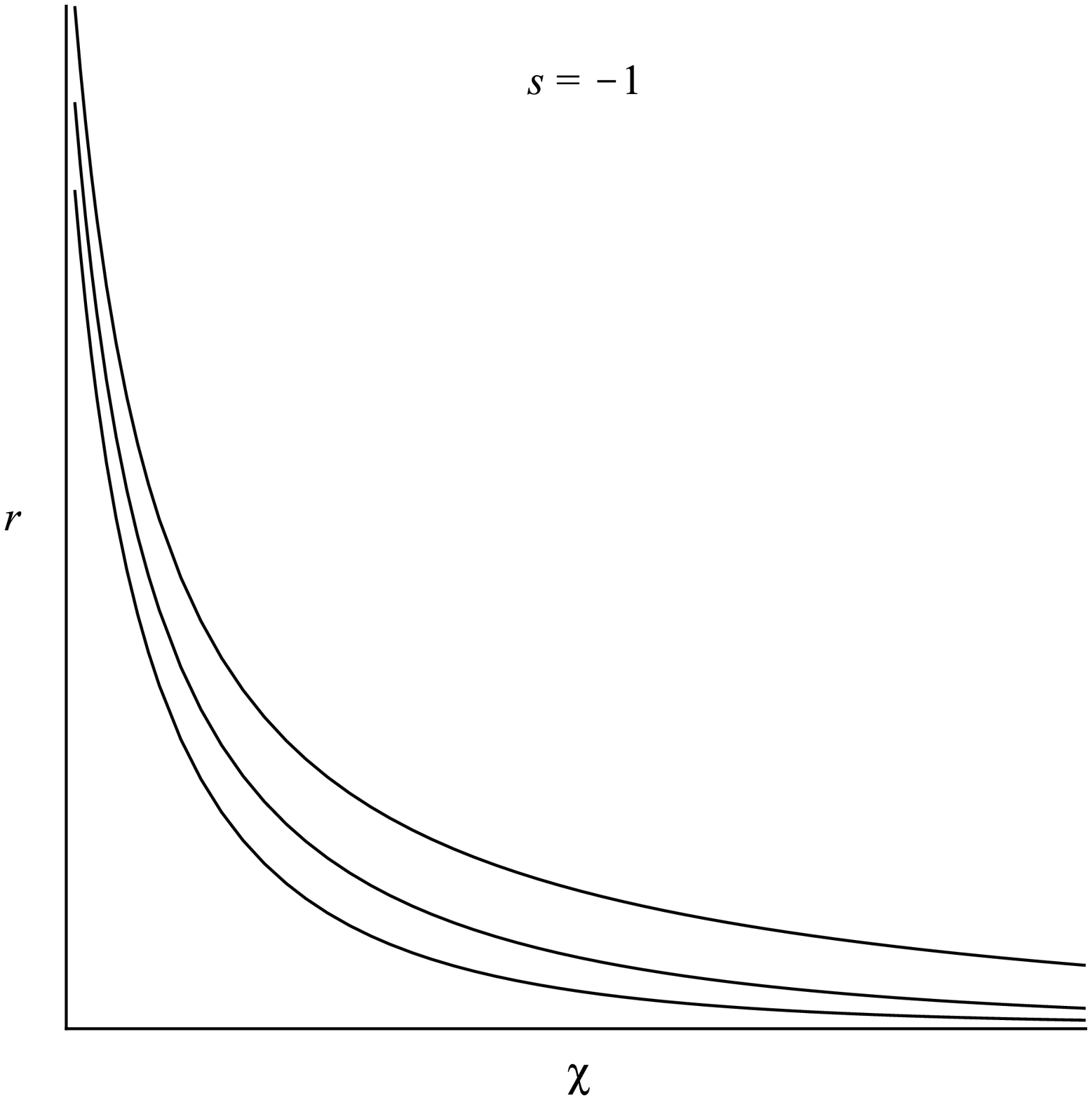}
 \caption{\label{fig:rdechialfamenor} \small Left: $r$ as a function of $\chi$ with $s = +1$ for the three possibilities of $\alpha$ with respect to $\alpha = 0$ in the $\alpha < 2$ case. $r$ has a critical point only in the case $0 < \alpha < 2$. For $\alpha = 0$ $r$ asymptotes to $r = R_S$ whereas for $\alpha < 0$ asymptotes to $r = 0$. Right: Case $s = -1$ for the same three possibilities of $\alpha$. In this case $r$ decreases monotonically towards $r = 0$ without critical points for all $\alpha < 2$.}
 \end{center}
\end{figure}

Now we study the full geometry of the solutions (\ref{metricalphalower2}).
In Fig.~\ref{fig:rdechialfamenor} we plot $r$ against $\chi$ according to (\ref{coordinatetransf}) and considering the two possibilities of $s$. We see that, in all cases, as $\chi$ departs from zero $r$ decreases from infinity; but in the $s = +1$ case $r$ may either reach a critical point for finite $\chi$ or be monotonically decreasing. From (\ref{coordinatetransf}) we find that the critical point is just the joining point $\hat{\chi}$ defined in (\ref{throat}), which only arises in the range $0 < \alpha < 2$ for the $s = +1$ solution. The other ranges of $\alpha$ in the $s = +1$ case give a monotonically decreasing $r$ with an asymptote for $\chi \rightarrow \infty$. It can be checked from Eq.~(\ref{coordinatetransf}) that the asymptotes are different for $\alpha = 0$ and $\alpha < 0$. We see again that the qualitative features of the solution $s= +1$ are discontinuously different in the ranges $\alpha = 0$, $0 < \alpha < 2$ and $\alpha < 0$. In the following we describe these solutions separately and also the $s=-1$ solution, for which $r$ is everywhere monotonically decreasing in $\chi$ towards the origin $r=0$.

\begin{itemize}
\item {\bf Schwarzschild solution for the $\alpha = 0$ point}\\
For $\alpha = 0$ the coordinate transformation (\ref{coordinatetransf}) can be explicitly inverted, yielding
\begin{equation}
 e^{-2 s \chi} = 1 - s \frac{2 C}{r} = N^2 = f \,.
\end{equation}
This is the Schwarzschild solution, as expected. The case $s=+1$ represents the positive-mass solution, whereas $s=-1$ is the negative-mass one. For the positive-mass case the radius $r$ asymptotes to the Schwarzschild radius $R_S = 2C$ as $\chi \rightarrow \infty$, as can be seen in Fig.~\ref{fig:rdechialfamenor}. Thus, (\ref{metricalphalower2}) with $s = +1$ in the $\alpha = 0$ case describes the exterior region of the (positive mass) Schwarzschild space-time. For $s= -1$ the coordinate $\chi$ covers the space up to the origin $r=0$, which in this case correspond to $\chi = +\infty$, and at the origin both $N^2$ and $f$ diverge.

\item {\bf $s=+1$ solution of the range $0 < \alpha < 2$}\\
The joining point $\hat{\chi}$ corresponds to a lower bound for $r$, which is equal to
\begin{equation}
 \hat{r} = 
 C \left( 
   \frac{ ( 1 - \beta^2 )^{(1+\beta) / 2}}{1 - \beta} 
   \right)^{1/\beta} \,.
\label{rhat}
\end{equation}
As $\chi$ continues beyond $\hat{\chi}$ the space extends itself increasing again the values of the radius $r$, as can be seen in Fig.~\ref{fig:rdechialfamenor}. This kind of space is a wormhole geometry constituted by two spatial branches joined by a throat, which is located at $\hat{\chi}$. This geometry was described by the authors of Ref.~\cite{Eling:2006df} as a solution of the EA theory. At the throat a 2-sphere of minimal area is reached. The spatial infinity we identified above with the value $\chi = 0$ and where we imposed the boundary conditions and obtained the asymptotic flatness corresponds to the infinite end of one of the branches. Let us call branch I to this sector. In the other branch, which we call branch II, $r$ also goes from $\hat{r}$ to the spatial infinity $r = \infty$ (This was called the ``interior'' branch in \cite{Eling:2006df}). The original radial coordinate $r$ can be used to cover separately the two branches from their respective infinite boundaries down to the throat $\hat{r}$, but, unlike $\chi$, it fails to cover the throat itself. The $\chi$ coordinate covers the throat and an open set around it that extends itself over the whole wormhole except at the infinite boundary. 

Notice that the function $f$ vanishes at the throat. This is just a coordinate singularity since $f$ is a metric component under the $r$ coordinate. It can be seen from (\ref{metricalphalower2}) that the metric components in the $\chi$ coordinate are all regular at $\hat{\chi}$. Actually $-N^2$ is one of these components; its value at the throat is given by $-e^{-2 \hat{\chi}/\beta}$, which approaches zero as $\alpha \rightarrow 0$. Moreover, the field equations (\ref{css2} - \ref{eom2}) themselves and the algebraic equation (\ref{bnegative}) cannot be trusted at the throat since these equations are written in the $r$ coordinate. It can be checked that the field equations in the $\chi$ coordinate are exactly solved, even at the throat. We may also see that the Ricci scalar (\ref{ricciscalar1}) is regular at the throat $\hat{\chi}$.

At the spatial infinity of the branch II ($\chi \rightarrow \infty$) there also arise singularities. But these are very different since in this case both $N^2$ and $f^{-1}$ vanish. This singularity was characterized in Ref.~\cite{Eling:2006df} in the context of the EA theory. In particular it was studied the finiteness of radial light rays directed to this boundary. However, in Ho\v{r}ava theory light rays do not necessarily represent an upper bound for matter velocity. At the infinity of branch II $^{(4)}R$ diverges if $0 < \alpha < 3/2$ and vanishes if $3/2 < \alpha <2$. Indeed, $^{(4)}R$ has a critical point somewhere in the wormhole when the range is $3/2 < \alpha < 2$; whereas it is monotonically decreasing from the zero value at the boundary of branch I to a negative divergence at the boundary of branch II when $0 < \alpha < 3/2$. To clearly contrast this, in Fig.~\ref{fig:ricci} we plot the Ricci scalar of the wormhole for these two cases. The critical point of the Ricci scalar is reached at
\begin{equation}
 \tanh{\chi_c} = 2\beta \,,
\end{equation}
which is bigger than $\hat{\chi}$. That is, the critical point of $^{(4)}R$ is reached beyond the throat, in branch II.

\begin{figure}
 \begin{center}
 \includegraphics[scale=0.40]{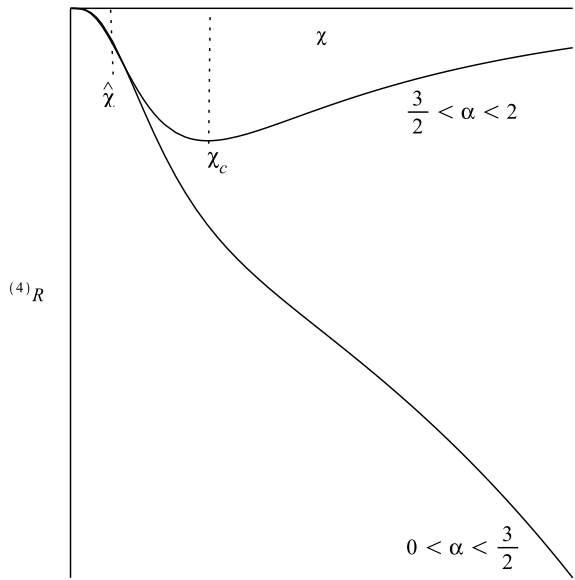}
  \caption{\label{fig:ricci} \small The four-dimensional Ricci scalar on the whole wormhole solution of the range $0 < \alpha <2$. For $0 < \alpha < 3/2$ it decreases monotonically towards a negative divergence at the boundary of branch II. For $3/2 < \alpha < 2$ it has a critical point after passing the throat and reaches again a zero at the boundary of branch II.}
 \end{center}
\end{figure}

For small $\alpha$, $\beta \rightarrow 1^-$, such that $\hat{\chi}$ tends to infinity. Therefore, the location $\hat{r}$ of the throat for $\alpha$ near to zero approximates to the Schwarzschild radius from above. We shall further develop on this connection in Section \ref{sec:perturbations}. The perturbative approach for small $\alpha$ we shall perform there yields the perturbative version of branch I of this exact solution.

The location $\hat{r}$ of the throat has also an upper bound in its running with respect to $\alpha$. As $\alpha$ approaches $\alpha = 2$, $\hat{r} \rightarrow e C$ from below ($e$ stands for the Euler number); that is, $\hat{r}$ becomes $e/2$ times the Schwarzschild radius. In general, $\hat{r}$ is monotonically increasing from $2C$ to $eC$ when run in $\alpha \in [0,2]$. We shall see that there is an interesting connection of this upper bound with the $\alpha > 2$ case.

\item {\bf $s=+1$ solution of the $\alpha < 0$ range}\\
The case $\alpha < 0$ for the $s=+1$ solution (\ref{metricalphalower2}) is qualitatively quite different to the previous two cases. The coordinate transformation (\ref{coordinatetransf}) is completely bijective in the full range of $\chi$. $r$ decreases monotonically from $r = \infty$ down to the origin $r=0$ as $\chi$ grows. Thus, there is no analogous for the throat of the $0 < \alpha < 2$ case nor horizon as in the Schwarzschild solution. The discontinuous lacking of a radial lower bound as one passes to the $\alpha < 0$ sector produces a failure in the perturbative solution for the $\alpha < 0$ case (see Section \ref{sec:perturbations}). At the point $\coth{\chi} = \beta$ the function $f$ has a critical point and after it increases monotonically towards $r = 0$. Thus, function $f$ is nonzero and regular in the whole domain of the radial coordinate $r$, except at $r = 0$ where it diverges. This divergence is a naked essential singularity (in the terminology appropiated for 
relativistic matter).\footnote{There is a proposal to distinguish from an observational point of view between naked singularities and a black holes \cite{Virbhadra:2002ju}.} The four-dimensional Ricci scalar $^{(4)}R$ (\ref{ricciscalar1}) is monotonically increasing from the zero value at $r = \infty$ towards a divergence at the origin $r=0$.
 
\item {\bf $s=-1$ solution, full range $\alpha < 2$}\\
This is the case of the asymptotically flat solution with negative mass. In Fig.~\ref{fig:rdechialfamenor} we plot $r$ as a function of $\chi$ for this case. The coordinate transformation behaves in a similar way to the $s=+1$ solution in the $\alpha < 0$ range. The transformation is bijective in its full range, $r$ decreases monotonically from $r=\infty$ to the origin $r=0$ as $\chi$ goes from zero to infinity. There is no lower bound of validity for the radial coordinate $r$, so the whole space can be covered both with it or $\chi$. Unlike the $s=+1$ solution, there are no subranges of $\alpha$ in which the behavior of the solution changes qualitatively; hence the physical features of the solution vary smoothly in the full range $-\infty < \alpha <2$. This includes the $\alpha = 0$ case, which is the negative-mass Schwarzschild solution. The metric is singular at the origin in both coordinate systems, where there is a naked essential singularity. The Ricci scalar (\ref{ricciscalar1}) decreases monotonically from the flat spatial infinite towards a negative divergence at the origin, which is labeled by $\chi = \infty$.  

\end{itemize} 


\subsection{Case $\alpha > 2$}
We take the following four solutions of Eq.~(\ref{bpositive}),
\begin{equation}
  \frac{\beta C}{r N} = s_1 \sin{\varphi} \,,
  \hspace{2em}
 \sqrt{f} + \frac{C}{r N} = s_2 \cos{\varphi} \,.
\label{solution2}
\end{equation}
In this case $\varphi$ arises as the appropriated transformed radial coordinate. There are four more solutions of Eq.~(\ref{bpositive}), they have the sine and cosine functions exchanged with respect to (\ref{solution2}). It turns out that those solutions lead to configurations that are equivalent to the ones (\ref{solution2}) yields, see Appendix \ref{apendix:equivalences}. Thus, solutions (\ref{solution2}) exhaust all the different solutions of the field equations.

The analysis of the domains of validity and joining of the local solutions in (\ref{solution2}) follows the same lines of the $\alpha < 2$ case. Positiveness of $r$ and $N$ leaves us with the domain $\varphi\in (0,\pi)$; in Appendix \ref{apendix:equivalences} we show that the domain $\varphi \in (\pi,2\pi)$ leads to the same solutions after a coordinate transformation. In $\varphi\in (0,\pi)$ positiveness of $\sqrt{f}$ leads to the domain of validity
\begin{enumerate}[i]
\item $s_1 = +1$, $s_2 = +1$ \,:\hspace{1em} solution in $\varphi\in (0,\arctan{\beta})$ with $C>0$.

\item $s_1 = +1$, $s_2 = -1$ \,:\hspace{1em} solution in $\varphi\in (\arctan{(-\beta)},\pi)$ with $C>0$.

\item $s_1 = -1$, $s_2 = +1$ \,:\hspace{1em} solution in $\varphi\in (0,\arctan{(-\beta)})$ with $C<0$.

\item $s_1 = -1$, $s_2 = -1$ \,:\hspace{1em} solution in $\varphi\in (\arctan{\beta},\pi)$ with $C<0$.
\end{enumerate}

In the $\alpha > 2$ case $\beta = \sqrt{\alpha/2 -1}$ and it can take any positive value, $\beta > 0$. Therefore, all of the four local solutions have nonempty domains of validity for any $\alpha > 2$. There are two global solutions, each one identified by the relative sign $s \equiv s_1 s_2$. Solutions (i) and (iv) are joined to form the $s=+1$ solution whereas (ii) and (iii) form the $s=-1$ solution. It turns out that the global $s=-1$ solution is equal to the $s=+1$ one after a coordinate change. We show this in Appendix \ref{apendix:equivalences}. Therefore, in the $\alpha > 2$ case there is only one solution. The joining point for solutions (i) and (iv) is given by
\begin{equation}
 \tan{\hat{\varphi}} = \beta \,.
\end{equation}

By repeating the same steps of the previous case, for the $s=+1$ solution we find the equations for the coordinates,
\begin{equation}
 \left( \frac{\cos{\varphi}}{\sin{\varphi}} - \frac{1}{\beta} \right)
  \frac{d\varphi}{dr} =
 - \frac{1}{r} \,,
\end{equation}
whose integral is
\begin{equation}
 r =
 \frac{k e^{\varphi / \beta}}{\sin{\varphi}} \,,
\label{coordinatetrans2}
\end{equation}
where $k > 0$. The value $\varphi = 0$ corresponds to the spatial infinity $r=\infty$. For any $\alpha > 2$ transformation (\ref{coordinatetrans2}) has a finite critical point exactly at the joining point $\hat{\varphi}$. This signals that this solution is very similar to the $s=+1$ solution of the $0 < \alpha < 2$ case, it is a wormhole.

The global expression of the solution in $\varphi\in (0,\pi)$ is
\begin{equation}
 N = \frac{\beta C}{k} e^{-\varphi / \beta} \,,
\hspace{2em}
 f = ( \beta^{-1} \sin{\varphi} - \cos{\varphi} )^2 \,,
\label{fngreaterthan2}
\end{equation}
where $C > 0$. Condition $N|_{\varphi = 0} = 1$ fixes $k = \beta C$. We can again write the metric tensor in terms of the $\varphi$ coordinate. It is 
\begin{equation}
 ds_{(4)}^2 =
 - e^{-2\varphi / \beta} dt^2
 + \frac{(\beta C)^2 e^{2\varphi / \beta}}{\sin^4{\varphi}}
    (d\varphi^2 + \sin^2{\varphi} d\Omega_{(2)}^2) \,.
\label{solutionalphagreater}
\end{equation}
The solution is valid in $\varphi \in (0, \pi)$. In this case the coordinate $\varphi$ allows to realize explicitly the local conformal equivalence between the solution and the 3-sphere with its standard metric in 3 spherical coordinates. The nonzero components of the Riemann tensor and the Ricci scalar are
\begin{eqnarray}
&&\begin{array}{l}
^{(4)}\!R_{t \varphi \varphi}{}^{t} =
 - 2 (\beta \sin{\varphi})^{-1} (\beta^{-1} \sin{\varphi} - \cos{\varphi}) \,,
\\[2ex]
^{(4)}\!R_{t \theta \theta}{}^{t} = 
 - {\displaystyle\frac{1}{2}} \sin^2{\varphi} R_{t \varphi \varphi}{}^{t} \,,
\hspace{2em}
^{(4)}\!R_{t \phi \phi}{}^{t} = \sin^2{\theta}\: {}^{(4)}\!R_{t \theta \theta}{}^{t} \,, 
\\[2ex]
^{(4)}\!R_{\varphi\theta\theta}{}^{\varphi} =
- \beta^{-1} \sin{\varphi} (\beta \sin{\varphi} + \cos{\varphi}) \,,
\hspace{2em}
^{(4)}\!R_{\varphi \phi \phi}{}^{\varphi} = 
 \sin^2{\theta}\:{}^{(4)}\!R_{\varphi\theta\theta}{}^{\varphi}  \,, \\[2ex]
^{(4)}\!R_{\theta \phi \theta}{}^{\phi} =
\beta^{-2}( 1 - (\beta \sin{\varphi} + \cos{\varphi})^2 ) \,,
\end{array}
\\ 
&&^{(4)}\!R = R =
 - {\displaystyle\frac{\alpha}{\beta^4 C^2}} e^{-2 \varphi / \beta} \sin^4{\varphi} \,.
\label{ricciscalar2}
\end{eqnarray}

Solution (\ref{solutionalphagreater}) also describes a wormhole formed by two spatial branches joined by a throat at $\varphi = \hat{\varphi}$, where a minimal 2-sphere is reached. The coordinate $\varphi$ covers the throat and a open neighborhood of it that extends itself up to infinity. By doing similar considerations as in the previous case, we get that at the spatial infinity labeled by $\varphi = 0$ the solution is asymptotically flat with positive mass,
\begin{equation}
  N^2 = f = 1 - \frac{2 C}{r} \,.
\end{equation}
The Ricci scalar (\ref{ricciscalar2}) vanishes at the infinity $\varphi = 0$. There is, however, an important difference with respect to the wormhole solution of the $0 < \alpha < 2$ range. At the other infinity $r=\infty$, which is labeled by $\varphi = \pi$, neither $N^2$ nor $f^{-1}$ vanish. On the contrary, they exhibit asymptotically flat behavior with negative mass,
\begin{equation}
 f = 1 + \frac{ 2 C e^{\pi / \beta}}{r} \,,
 \hspace{2em} 
 N^2 = e^{- 2 \pi / \beta} f \,.
\end{equation}
Consequently, Ricci scalar (\ref{ricciscalar2}) also vanishes at this end.
Actually, both in the $0 < \alpha < 2$ and $\alpha > 2$ wormholes $N^2$ is monotonically decreasing as one walks from the infinite boundary of one branch to the infinite boundary of the other one, including the passing through the throat. But the difference arises when the $\varphi$ coordinate in the $\alpha > 2$ case reaches its upper bound, $\varphi = \pi$. There the function $N^2$ ends with a finite nonzero value. In the wormhole of the $0 < \alpha < 2$ case the $\chi$ coordinate runs up to infinity and the function $N^2$ decreases completely to zero.

Thus, we have that both branches of this wormhole solution exhibit asymptotically flat behavior. In one branch the maximum value of $N^2$ is reached at the infinite boundary whereas in the other branch the opposite case occurs: at the infinite boundary the minimum (nonzero) of $N^2$ is reached. This renders as a rather artificial matter the choice of one or the other branch as the reference place to demand asymptotic flatness and hence indicating the sign of the mass of a plausible external source.

As in the wormhole of the $0 < \alpha < 2$ range, function $f$ vanishes at the throat; it can be checked that the field equations are consistently solved once they are written in the $\varphi$ coordinate. Four-dimensional Ricci scalar is regular at the throat and has a critical point at $\tan{\varphi_c} = 2\beta$, which again is a location bigger than the throat. Unlike the wormhole of the $0 < \alpha <2$ range, this solution is completely regular.

\begin{figure}[t]
 \begin{center}
 \includegraphics[scale=0.40]{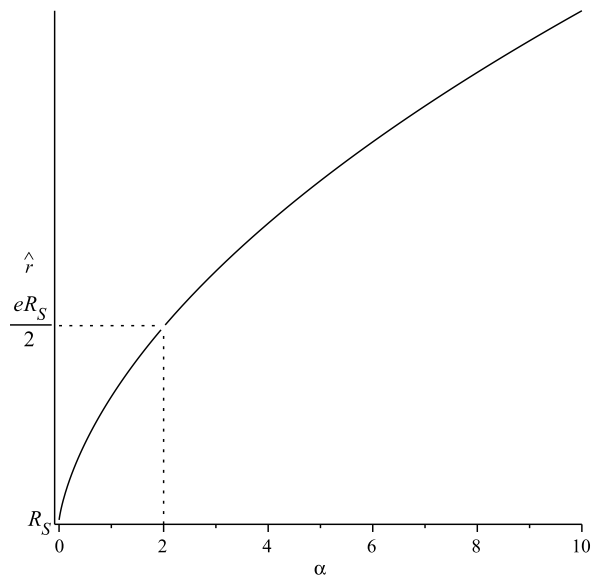}
 \caption{\label{fig:rhat} \small The location $\hat{r}$ of the throat of the wormhole solutions of the ranges $0 < \alpha <2$ and $\alpha > 2$ varied as a function of the coupling constant $\alpha$. For the $0 < \alpha < 2$ solution the running starts just above $R_S$ and ends just below $e/2$ times the Schwarzschild radius; whereas for the $\alpha > 2$ solution starts just above of that value.}
 \end{center}
\end{figure}

Interestingly, as $\alpha$ approaches $\alpha = 2$ from above, $\beta \rightarrow 0^+$, such that $\hat{\varphi} \rightarrow 0$ and $\hat{r} \rightarrow e C$ from above. Recall that in the $0 < \alpha < 2$ solution $\hat{r}$ goes to the same value from below when $\alpha \rightarrow 2^-$. Thus, the solution (\ref{solutionalphagreater}) can be regarded in some sense as the continuation of the $s=+1$ solution of the $0 < \alpha < 2$ case; although the continuation cannot be passed through the $\alpha = 2$ point, as we are going to see below, and the singularity is regularized at the end of branch II. The location $\hat{r}$ of the throat is monotonically increasing when run in $\alpha \in (2,\infty)$, spanning the range $(e C, \infty)$. In Fig.~\ref{fig:rhat} we put together the runnings of $\hat{r}$ with respect to $\alpha$ for the cases $0 < \alpha < 2$ and $\alpha > 2$.

\subsection{Case $\alpha = 2$} 
To analyze this special subspace of parameters we need to come back to the original static field equations (\ref{eisnteinstatic}) and (\ref{hamiltoniancstatic}). When $\alpha = 2$, Eq.~(\ref{hamiltoniancstatic}) is just the trace of Eq.~(\ref{eisnteinstatic}), thus in this case the theory lacks one field equation for static configurations. Eq.~(\ref{eisnteinstatic}) can be rewritten in the form
\begin{equation}
 R^{ij} - N^{-1} ( \nabla^i \nabla^j N + g^{ij} \nabla^2 N ) 
   + 2 N^{-2} \nabla^i N \nabla^j N = 0 \,.
\label{eomalpha2}
\end{equation}
After inserting the static spherically symmetric ansatz with vanishing shift function (\ref{ansatz}) into this equation, we get that it leads to two different equations:
\begin{eqnarray}
 \frac{N''}{N} + \frac{f' N'}{2 f N} + \frac{f'}{2 r f}
 + \frac{N'}{r N} - \left( \frac{N'}{N} \right)^2 &=& 0 \,,
\label{eom1alpha2}
\\
 \frac{N''}{N} + \frac{f' N'}{2 f N} + \frac{f'}{2 r f}
 + \frac{3 N'}{r N} + \frac{f - 1}{r^2 f} &=& 0 \,.
\label{eom2alpha2}
\end{eqnarray}
A combination of these two equations yields
\begin{equation}
 \left( \frac{r N'}{N} + 1 \right)^2 = \frac{1}{f} \,.
\end{equation}
This has the two solutions Kiritsis \cite{Kiritsis:2009vz} found: $N$ is given in terms of $f$ as
\begin{equation}
 \frac{r N'}{N} = -1 \pm \frac{1}{\sqrt{f}}
\label{nalpha2}
\end{equation}
and $f$ is arbitrary. It is straightforward to check that (\ref{nalpha2}) solves Eqs.~(\ref{eom1alpha2}) and (\ref{eom2alpha2}) for all $f(r)$. Therefore, we see that in the $\alpha = 2$ case the lacking of one field equation leaves one metric function indeterminate. Because of this, the $\alpha = 2$ case was called degenerated by Kiritsis.


\section{Perturbative solutions}
\label{sec:perturbations}
In this section we proceed to solve the Eqs.~(\ref{css2} - \ref{eom2}) in an approximate way by assuming an small value of $\alpha$. We call such solutions the \emph{perturbative solutions}. Since the approximation consists of expanding the field equations up to linear order in $\alpha$, it is equivalent to do perturbations on the Schwarzschild solution with a scale of the order of $\alpha$ for the perturbations. For solving the field equations the strategy will consist of combining Eqs.~(\ref{css2}) and (\ref{eom2}) to obtain a differential equation for $f$ that can be solved perturbatively in $\alpha$. Then, we shall put the solution for $f$ back into Eq.~(\ref{eom2}) and manage to find the perturbative solution for $N$. Finally, we shall check that Eq.~(\ref{eom1}) is solved up to linear order by the $f$ and $N$ found.

One may solve Eq.~(\ref{eom2}) for $N'/N$ and substitute the resulting expression into Eq.~(\ref{css2}), obtaining an equation for $f$ that can be written in the form
\begin{equation}
 8 ( 1 - h ) (r h)' 
  + \alpha \left[ \left( (r h)' - h \right)^2
       + 4 (r h)' \right] = 0 \,,
\label{eqh}
\end{equation}
where
\begin{equation}
 h \equiv 1 - f \,.
\end{equation}

Now we start the perturbations. We assume that the functions $f$ and $N$ are linear-order polynomials in $\alpha$. At zeroth order in $\alpha$, Eq.~(\ref{eqh}) reduces to $(r h^{(0)})' = 0$; its solution is proportional to $r^{-1}$; hence, we obtain the Schwarzschild factor
\begin{equation}
 f^{(0)} = 1 - \frac{A}{r} \,,
\end{equation}
where $A$ is an integration constant. By substituting $f^{(0)}$ into Eq.~(\ref{eom2}) we obtain the zeroth-order lapse function
\begin{equation}
 N^{(0)} = \left( 1 - \frac{A}{r} \right)^{1/2} \,,
\end{equation}
where the multiplicative integration constant that arises in this step has been fixed to unity by imposing the boundary condition $N|_\infty = 1$.

The linear-order function $h$ is obtained by expanding $h = h^{(0)} + \alpha h^{(1)}$ and substituting this expansion and the solution for $h^{(0)}$ into Eq.~(\ref{eqh}). After expanding the resulting equation up to linear order in $\alpha$, we obtain an equation for $h^{(1)}$,
\begin{equation}
 (r h^{(1)})' = 
  - \frac{A^2}{8 r^2} \left( 1 - \frac{A}{r} \right)^{-1} \,,
\end{equation}
which can be integrated straightforwardly,
\begin{equation}
 h^{(1)} = 
 -\frac{A}{8 r} \ln{\left( 1 - \frac{A}{r} \right)} + \frac{B}{r}\,,
\end{equation}
where $B$ is an integration constant.

Similarly, to obtain the linear-order $N$ we expand $N = N^{(0)} + \alpha N^{(1)}$ and substitute into Eq.~(\ref{eom2}). By expanding the resulting equation up to linear order, we get
\begin{equation}
 \left( \frac{ N^{(1)} }{ N^{(0)} } \right)' =
 - \frac{1}{16 r^2} \left(1 - \frac{A}{r}\right)^{-2} 
  \left[ \frac{A^2}{r} 
        + A \ln{\left(1 - \frac{A}{r}\right)}  
 - 8 B \right] \,.
\end{equation}
Its integral is
\begin{equation}
 \frac{ N^{(1)} }{ N^{(0)} } =
  \frac{1}{8}\left( 1 - \frac{A}{r} \right)^{-1}\left[
    1 + \left( 1 - \frac{A}{2 r} \right) 
     \ln{\left( 1 - \frac{A}{r} \right)}
     - \frac{4 B}{A} \right]
 + D \,,
\end{equation}
where $D$  is an integration constant. The boundary condition $N|_\infty = 1$ fixes $D = ( 4 B / A - 1 )/8$; thus, we have
\begin{equation}
 N^{(1)} = 
  \frac{1}{8} \left( 1 - \frac{A}{r} \right)^{-1/2}
   \left[ \frac{A - 4 B}{r} + \left( 1 - \frac{A}{2 r} \right) 
     \ln{\left( 1 - \frac{A}{r} \right)} \right] \,.
\end{equation}
This exhausts the linearized Eqs.~(\ref{css2}) and (\ref{eom2}). It is a matter of straightforward computations to check that the solutions we have found for $f$ and $N$ solve the linearized version of Eq.~(\ref{eom1}).

We have arrived at the linear-order perturbative solution
\begin{equation}
\begin{array}{rcl}
 N(r) &=& {\displaystyle
  \left( 1 - \frac{A}{r} \right)^{1/2} 
  + \frac{\alpha}{8} \left( 1 - \frac{A}{r} \right)^{-1/2}
     \left[ \frac{A - 4 B}{r} + \left( 1 - \frac{A}{2 r} \right) 
     \ln{\left( 1 - \frac{A}{r} \right)} \right]  } \,,
\\[2ex]
 f(r) &=& {\displaystyle
  1 - \frac{A + \alpha B}{r} 
  + \frac{\alpha A}{8 r} \ln{\left( 1 - \frac{A}{r} \right)} } \,,
\end{array}
\label{perturbativesolution}
\end{equation}
which is valid for the range $r > A$. In the $\alpha = 0$ case this solution reproduces the Schwarzschild metric, $A$ being the Schwarzschild radius.

Unlike the Schwarzschild solution of general relativity which has only one arbitrary integration constant, the solution we have obtained has two integration constants: $A$ and $B$. This apparent excess of free parameters has no physical meaning nor it is a mathematical inconsistency; it is just a consequence of the fact that we are not dealing with exact solutions, but with approximated ones. One may check that, when evaluating the Eqs.~(\ref{css2} - \ref{eom2}) on the perturbative solution, the parameter $B$ gets involved only in terms of the quadratic or higher order in $\alpha$. Thus, there is no way to fix it by solving the field equations at linear order in $\alpha$. On the other hand, in Appendix \ref{appendix:expanding} we expand directly the exact solution up to linear in $\alpha$, obtaining the value $B = 0$. Hence, for now on we take this value for $B$.

We study the behavior of the perturbative solution (\ref{perturbativesolution}) at two limits: near the Schwarzschild radius and the asymptotic limit for large $r$. For simplicity, let us restrict the integration constant $A$ to be positive. This will leads us to the positive mass solutions, which are the most interesting ones physically (the perturbative solution (\ref{perturbativesolution}) is valid for both cases).

\begin{figure}[t]
\begin{center}  
 \includegraphics[scale=0.40]{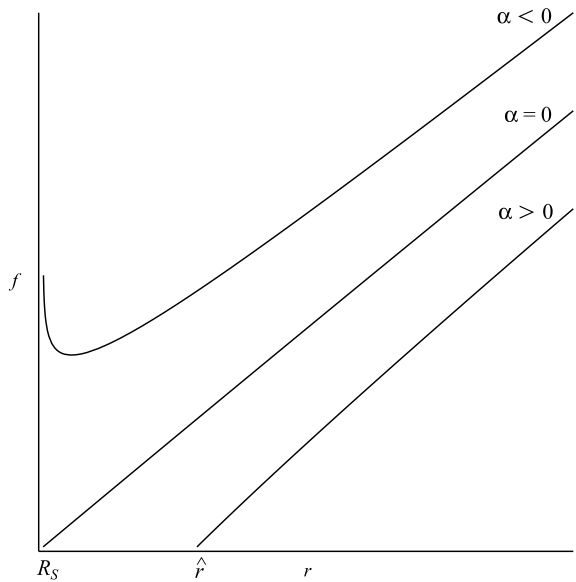} 
   \caption{\label{fig:perturbativefnearhorizon}\small The perturbative function $f$ near the Schwarzschild radius $R_S$ for the three cases of $\alpha$. The curve $\alpha = 0$ is the plot of the Schwarzschild factor which vanishes at $R_S$. The function $f$ for $\alpha > 0$ is monotonically decreasing towards the throat $\hat{r}$ where it vanishes. For $\alpha < 0$ $f$ extends its domain down to $R_S$. Near $R_S$ it reaches a minimum and after it increases monotonically towards a divergence at $R_S$.}
 \end{center}
\end{figure}

We start by studying the behavior of the $f$ function,
\begin{equation}
  f(r) =   
   1 - \frac{A}{r} 
  + \frac{\alpha A}{8 r} \ln{\left( 1 - \frac{A}{r} \right)}  \,,
\end{equation}
for $r \sim A = R_S$. In Fig.~\ref{fig:perturbativefnearhorizon} we plot the function $f(r)$ at this limit for the three possibilities of $\alpha$. We recall that the domain of validity of the solution falls into $0 < 1 - A/r < 1$ for all $\alpha$. If $\alpha > 0$ function $f$ has a root at some value $\hat{r}$ greater than and near the Schwarzschild radius, $\hat{r} \gtrsim A$, given by the solution of the equation\footnote{The limit of $\alpha \ln{\alpha}$ when $\alpha \rightarrow 0^+$ is finite and equal to zero.}
\begin{equation}
 1 - \frac{A}{\hat{r}} 
  = - \frac{\alpha A}{8 \hat{r}} \ln{\left( 1 - \frac{A}{\hat{r}} \right)} \,.
\label{rootf}
\end{equation}
Therefore, the perturbative solution is valid up to the value $\hat{r}$ where it has a (coordinate) singularity. This $\hat{r}$ is the perturbative version of the throat we found in the exact solution, which holds for $\alpha > 0$. 

On the other hand, in the case $\alpha < 0$ the function $f$ has not any root (Eq.~(\ref{rootf}) has no solution in $A < r < \infty$ with $\alpha < 0$); hence, its domain extends down to the Schwarzschild radius $A$. Instead, near the Schwarzschild radius it has a critical point given by the equation
\begin{equation}
 1 - \frac{A}{r} =
 \frac{\alpha}{8} \left[
   \left( 1 - \frac{A}{r} \right) \ln{\left( 1 - \frac{A}{r} \right)}
   - \frac{A}{r} \right] \,.
\end{equation}
As one moves from this critical point towards the Schwarzschild radius the function $f$ grows monotonically without upper bound, exhibiting a divergence at $r = A$. This behavior departs drastically from Schwarzschild solution, which always decreases monotonically towards the Schwarzschild radius. Since the difference between this two functions increases unavoidably as one approaches the Schwarzschild radius (in a region of nonzero measure), the perturbative solution cannot be trusted for the case $\alpha < 0$ if one is interested in the region near the Schwarzschild radius. For $\alpha > 0$ the perturbative solution is totally admissible in the range $\hat{r} > r > \infty$. $\hat{r}$ can get close to the Schwarzschild radius as wish by lowering $\alpha$. The sector $\hat{r} > r > \infty$ corresponds to the branch I of the exact wormhole solution. As in the exact solution, function $N^2$ decreases monotonically as one goes to the throat, reaching a nonzero value there. Notice that the $\alpha < 0$ perturbative solution can still be admissible for values of the radius much bigger than the Schwarzschild radius.

We now study the asymptotic behavior of the perturbative solution (\ref{perturbativesolution}). Notice that neither $N^2$ nor $f$ have contributions of order $\alpha$ to the mode $1/r$; indeed, this mode is $A/r$ for both functions, thus we identify the integration constant as $A = 2 G M$, where $M$ is the mass of an external source. The asymptotic expansion, up to $1/r^3$ order, of the perturbative solution is 
\begin{equation}
\begin{array}{rcl}
 N^2 &=& 
 {\displaystyle 1 - \frac{2 G M}{r} - \frac{\alpha (2 G M)^3}{48 r^3} 
 + \mathcal{O}\left(\frac{1}{r^4}\right) } \,,
\\[2ex]
 f &=& 
 {\displaystyle 1 - \frac{2 G M}{r} - \frac{\alpha (2 G M)^2}{8 r^2} - \frac{\alpha (2 G M)^3}{16 r^3} + \mathcal{O}\left(\frac{1}{r^4}\right) } \,.
\end{array}
\label{asymptoticexpansion}
\end{equation}

One may contrast this asymptotic expansion with the exact solution without need of finding the latter explicitly, since it is easy to check whether an expansion in $1/r$ as (\ref{asymptoticexpansion}) solves the field equations up to a given order in $1/r$. Remarkably, it turns out that (\ref{asymptoticexpansion}) is the asymptotic solution, up to $1/r^3$ order, of the field equations (\ref{css2} - \ref{eom2}) \emph{without expanding in $\alpha$}. That is, (\ref{asymptoticexpansion}) is precisely the asymptotic expansion of the \emph{exact} solution up to $1/r^3$ order, which turns out to be of linear order in $\alpha$. The asymptotic expansion we show in (\ref{asymptoticexpansion}) for the perturbative and exact solutions coincides with the expansion of the exact solution shown in Ref.~\cite{Kiritsis:2009vz}, except for the sign of the $1/r^3$ term in the $N^2$ expansion, which we found to be negative.


\section*{Discussion and conclusions}
We have obtained explicitly the static spherically symmetric solutions of the complete nonprojectable Ho\v{r}ava theory, which depends on the vector $\partial_i \ln N$. We have found the components of the space-time metrics explicitly as functions of local coordinates. We have found the solutions for the lowest-order effective action (without cosmological constant) since this kind of configurations are mainly interesting for large-distance physics. We have imposed the condition of vanishing of the shift function in order to simplify the computations. Configurations of the same kind but with nonzero shift function deserve to be further investigated. The only undetermined coupling constant the solutions have is the one of the $(\partial_i \ln N)^2$ term, which is $\alpha$. Indeed, although the $\lambda = 1/3$ and $\lambda \neq 1/3$ theories are in general qualitatively different since their number of propagating degrees of freedom differ (two for the former and three for the latter), the static solutions (with vanishing shift function) are the same for both cases since there is no influence of kinetic terms on such configurations. In Table 1 we show the several solutions found according to the ranges of $\alpha$. 

By managing the Lagrangian field equations in order to extract the constraints of the theory (see Appendix \ref{appendix:hamiltonian}), we have obtained an algebraic field equation for $N$ and $f$ that could be solved in a closed way. This equation splits out in the cases $\alpha < 2$ and $\alpha > 2$. Its solutions are given in terms of one-parameter families. In all cases we have regarded the free parameter as a transformed radial coordinate. With the transformed coordinate we have established explicit local conformal equivalences between the solutions and standard geometries. In the case $\alpha < 2$ we have ended up with hyperbolic coordinates whereas spherical coordinates arise in $\alpha > 2$. 

Although we have not restricted \emph{a priori} the range of $\alpha$ as would be required by the linear stability of an extra mode since it is absent in the $\lambda = 1/3$ case, the qualitative features of the solutions in the $\alpha < 2$ range differ discontinuously among the $\alpha < 0$ and $0 < \alpha < 2$ subranges, the point $\alpha = 0$ being the case of GR. In spite of this, with the hyperbolic coordinates the master expression for the solutions of these three cases can be given in an unified way.

\begin{table}[t]
\caption{\label{tableofsolutions} \small The several kinds of static spherically symmetric solutions of the complete nonprojectable Ho\v{r}ava theory with vanishing shift function classified according to the values of the coupling constant $\alpha$. We indicate the number of different solutions found in each range. The singularities are the essential ones. All solutions except $\alpha =2$ are unique up to a physical integration constant.}
\begin{center}
\begin{tabular}{| c  c  l  l  l |}
\hline 
{\bf Range} & \multicolumn{1}{c}{\bf Number} & 
\multicolumn{1}{c}{\bf Geometry} & \multicolumn{1}{c}{\bf Mass} & 
\multicolumn{1}{c|}{\bf Singularities} 
  \\ \hline
\multirow{2}{*}{$\alpha <  0$} & \multirow{2}{*}{2} & 
1. Naked sing. & Positive & At the origin \\
 & & 2. Naked sing. & Negative &  At the origin \\
\hline
\multirow{2}{*}{$\alpha = 0$} & \multirow{2}{*}{2} & 
1. Schwarzschild & Positive & At the interior \\
 & & 2. Schwarzschild & Negative & At the origin \\
\hline
\multirow{2}{*}{$0 < \alpha < 2$} & \multirow{2}{*}{2} &
1. Wormhole & Positive & Boundary of branch II \\
 & &  2. Naked sing. & Negative & At the origin \\ 
\hline
$\alpha = 2$ & 
\multicolumn{4}{c|}{Degenerate case: solutions have a hair. }  \\
\hline
$\alpha > 2$ & 1 & Wormhole & 
Undefined & Completely regular \\
 \hline
 \end{tabular}
\end{center}
\end{table}

In the ranges $0 < \alpha < 2$ and $\alpha > 2$ there arise wormhole solutions. The coordinate systems we have used are valid at each throat and in a open neighborhood around them. Such carts extend themselves over the whole branches except at infinity.

All solutions (except the degenerated case) are asymptotically flat at least in one sector of the solution. In the wormhole of the $0 < \alpha < 2$ range the asymptotic flatness is manifested with positive mass at the end of branch I, whereas there is a singularity at the end of branch II. Curiously, the wormhole of the $\alpha > 2$ range is asymptotically flat at the ends of its two branches,  but, due to the monotonic decreasing of the lapse function over the whole wormhole, the signs of the corresponding masses are different for each branch.

For $\alpha$ near zero and positive, we have that branch I of the wormhole solution tends smoothly to the exterior region of the (positive mass) Schwarzschild space-time.  In particular, the location of the throat tends to the Schwarzschild radius for $\alpha \rightarrow 0^+$. Moreover, for locations sufficiently above the Schwarzschild radius, even the negative, small-$\alpha$ solution with positive mass is a small deformation of the exterior Schwarzschild space-time. These results are important for the coupling to stellar matter, where there arise cutoffs for the radial validity of the vacuum solutions. In these scenarios vacuum solutions, as the ones studied here, are of interest only as exterior solutions. In this sense it is interesting to note that the perturbative solution we found is a smooth deformation of the exterior region of the Schwarzschild space-time written directly in the original radial coordinate $r$.

An interesting extension of our work would be the inclusion of a cosmological constant. It is plausible that coordinate transformations similar to the ones we performed here work as well for the case of the field equations of the large-distance effective Ho\v{r}ava theory with a nonvanishing cosmological constant evaluated on static spherically symmetric configurations. In particular, one may elucidate whether there is a minimum for the coordinate transformation, which would signal the presence of a wormhole. We expect to report on this shortly.

We have studied the solutions in the framework of the effective action of the complete nonprojectable Ho\v{r}ava theory. We may briefly compare these solutions with other developments of the original Ho\v{r}ava proposal. All the projectable versions, among them the $f(R)$ models of Refs.~\cite{Kluson:2009xx,Chaichian:2010yi}, automatically exclude these solutions since their lapse function has a nontrivial dependence on the radial coordinate, and there is no allowed coordinate change that can absorb this dependence. On the nonprojectable side, any truncated model allowing the $\partial_i \ln{N}$ terms of Blas, Pujol\`as and Sibiryakov \cite{Blas:2009qj} should possess these solutions as their large-distance approximation of the static spherically symmetric solutions (with vanishing $N_i$), since we are studying the most general effective action (those that do not include the $\partial_i\ln{N}$ terms, as the ones with detailed balance, find the Schwarzschild space-time as their large-distance limit within the space of static spherically symmetric configurations). There is a version of the Ho\v{r}ava theory with a further $U(1)$ gauge symmetry originally proposed for the projectable case \cite{Horava:2010zj} and later extended to the nonprojectable one \cite{Zhu:2011xe}. In vacuum, which is the case of the solutions studied here, both of these models have the constraint $^{(3)}R = \mbox{constant}$ forced by the presence of an additional gauge field. This highly restrictive constraint excludes the possibility of having spatial slices of nonconstant curvature, as the wormholes or the naked singularities we have found here.

Finally, we point out that there are more static spherically symmetric solutions once we discard the restriction of vanishing shift function. In particular, it has been shown numerically that static spherically symmetric black holes exist in the EA theory \cite{Barausse:2011pu,Eling:2006ec}. They have the aether vector field with both timelike and spacelike components turned on. These black holes must also arise in the the large-distance effective action of the complete Ho\v{r}ava theory since both theories are physically equivalent. In this case the presence of spacelike components of the aether vector field must be equivalent to the activation of the shift function on the side of the Ho\v{r}ava theory.


\section*{Acknowledgments}
A. R. and A. S. are partially supported by Project Fondecyt No.
1121103, Chile.


\appendix
\section{The solution in the Hamiltonian formalism}
\setcounter{section}{1}\setcounter{equation}{0}
\label{appendix:hamiltonian}
We focus the solutions in the Hamiltonian formulation of the $\lambda = 1/3$ theory. After this, we shall comment on the $\lambda \neq 1/3$ case.

The bulk part of the Hamiltonian of the complete nonprojectable theory at $\lambda =1/3$ is a sum of local constraints \cite{Bellorin:2013zbp},
\begin{eqnarray}
 H = \int d^3x \left(
       N \mathcal{H} + N_i \mathcal{H}^i + \sigma \phi + \mu \pi 
           \right)  \,.
\label{hamiltonianfinal}
\end{eqnarray}
The shift $N_i$ as well as $\sigma$ and $\mu$ play the role of Lagrange multipliers. The first-class constraint is the momentum constraint $\mathcal{H}^i \equiv - 2 \nabla_j \pi^{ij} + \phi \partial^i N = 0$. The second class ones are $\phi = 0$, $\pi =0$, the Hamiltonian constraint $\mathcal{H} = 0$ and $\mathcal{C}=0$, where
\begin{eqnarray}
 \mathcal{H} &\equiv&
 \frac{1}{\sqrt{g}} \pi^{ij} \pi_{ij} + \sqrt{g} \tilde{\mathcal{V}} \,,
\label{hamiltonianconstraintgeneral}
\\
\mathcal{C} &\equiv&
\frac{3N}{2\sqrt{g}} \pi^{ij} \pi_{ij} 
- \sqrt{g} \tilde{\mathcal{V}}\,' \,,
\label{cconstraint}
\end{eqnarray}
and we have introduced the modified potential and its derivative
\begin{eqnarray}
\tilde{\mathcal{V}} & \equiv &
\mathcal{V} 
   + \frac{1}{N} \sum\limits_{r=1} (-1)^r 
      \nabla_{i_1 \cdots i_r} \left( N 
         \frac{\partial \mathcal{V}}{\partial ( \nabla_{i_r \cdots i_2} a_{i_1} )} 
          \right) \,,
\label{modifiedpotential}
\\
 \tilde{\mathcal{V}}\,' &\equiv& 
  \frac{1}{\sqrt{g}} g_{ij} \frac{\delta}{\delta g_{ij}} \int d^3y \sqrt{g} N \tilde{\mathcal{V}} \,.
\label{vprima}
\end{eqnarray}

For the large-distance effective action we have $\mathcal{V} = \mathcal{V}^{(2)}$, such that the Hamiltonian and $\mathcal{C}$ constraints become
\begin{eqnarray}
 \frac{1}{\sqrt{g}} \mathcal{H} & = &
  \frac{1}{g} \pi^{ij} \pi_{ij} - R 
  + 2 \alpha N^{-1} \nabla^2 N - \alpha a_i a^i  \,,
\label{hamiltonianconstrainquadratic}
\\
 \frac{1}{\sqrt{g} N} \mathcal{C} & = &
 \frac{3}{2 g} \pi^{ij} \pi_{ij}
 + \frac{1}{2} R - 2 N^{-1} \nabla^2 N
 + \frac{\alpha}{2} a_i a^i \,.
\label{cconstraintquadratic}
\end{eqnarray}
The system $\mathcal{H}=0$, $\mathcal{C}=0$ can be brought to the form
\begin{eqnarray}
 g^{-1} \pi^{ij} \pi_{ij} 
   + ( \alpha / 2 - 1 ) N^{-1} \nabla^2 N 
  & = & 0 \,,
\label{c1}
\\
 R - ( 1 + 3 \alpha / 2 ) N^{-1} \nabla^2 N 
  + \alpha a_i a^i & = & 0  \,.
\label{c2}
\end{eqnarray}
The preservation in time of the second-class constraints leads to a system of two equations for the Lagrange multiplier $\sigma$ and $\mu$. This system is
\begin{eqnarray}
  \beta \left( 2 \nabla^2 \sigma + N a^i \partial_i \mu \right)
- 2 g^{-1} \pi^{ij} \pi_{ij} \sigma
+ \left( \beta \nabla^2 N + 3 g^{-1} N \pi^{ij} \pi_{ij} \right) \mu
= &&
\nonumber \\
 - \frac{4 N}{\sqrt{g}} \pi^{ij} ( N R_{ij} -\nabla_i \nabla_j N  
   + \alpha N a_i a_j ) 
 + \frac{4\beta}{\sqrt{g}} \partial_i ( N \partial_j N \pi^{ij} ) 
\,, &&  
\label{sigmaeq}
\\
  \nabla^2 \mu 
- \frac{\alpha}{N} a^i \partial_i \sigma
- \frac{1}{4} \left(
   R + \alpha a_i a^i + (3/\gamma) g^{-1} \pi^{ij} \pi_{ij} \right) \mu
+ \frac{\alpha}{N} a_i a^i \sigma  
= \,\,\, \hspace{2em} &&
\nonumber \\
 \frac{2\alpha}{\beta \sqrt{g}} \pi^{ij} \left( N R_{ij} 
     - \nabla_i \nabla_j N + \alpha N a_i a_j \right) 
\,, &&
\label{mueq}
\end{eqnarray}
where
\begin{equation}
\beta \equiv ( 1 - \alpha/2 ) \,,
\hspace{2em}
\gamma \equiv \left( \frac{ 1 - \alpha/2 }{ 1 + 3\alpha/2} \right) \,.
\end{equation}

Now we move to the canonical equations of motion for the $\lambda = 1/3$ theory. Since $\phi = 0$ is a constraint of the theory, $\dot{\phi}$ vanishes in the totally constrained phase space with no more conditions on the canonical variables. The equations for the evolution of $N$ and $g_{ij}$ are
\begin{eqnarray}
 \dot{g}_{ij} &=& 
 \frac{2 N}{\sqrt{g}} \pi_{ij} + 2 \nabla_{(i} N_{j)}
 + \mu g_{ij} \,,
\label{dotg}
\\
 \dot{N} &=& \sigma + N^k \nabla_k N \,.
\label{dotn}
\end{eqnarray}
Equation (\ref{dotg}) and the constraint $\pi = 0$ imply the relation
\begin{equation}
 g^{kl} \dot{g}_{kl} = 2 \nabla_k N^k + 3 \mu \,.
\label{tracedotg}
\end{equation}
The last equation of motion is
\begin{equation}
\begin{array}{rcl}
  \dot{\pi}^{ij} &=& {\displaystyle
  -\frac{2 N}{\sqrt{g}} ( \pi^{ik} \pi_k{}^j 
    -\frac{1}{4} g^{ij} \pi^{kl} \pi_{kl} )
  - \sqrt{g} N ( R^{ij} - \frac{1}{2} g^{ij} R ) }
\\[2ex] & & {\displaystyle
  + \sqrt{g} ( \nabla^i \nabla^j N - g^{ij} \nabla^2 N )
  - \alpha \sqrt{g} N ( a^i a^j 
     - \frac{1}{2} g^{ij} a_k a^k ) }
\\[2ex] & &  
  - 2 \nabla_k N^{(i} \pi^{j)k}
  + \nabla_k ( N^k \pi^{ij})
  - \mu \pi^{ij}  \,.
\end{array}
\label{dotpi}
\end{equation}

Let us evaluate all the equations of motion and constraints for static configurations with vanishing shift function. From Eqs.~(\ref{dotn}) and (\ref{tracedotg}) we get that the Lagrange multipliers $\sigma$ and $\mu$ vanish. Putting this information back into Eq.~(\ref{dotg}) yields that static configurations with vanishing shift function necessarily have vanishing canonical momentum, $\pi^{ij}= 0$. This automatically solves the $\pi = 0$ and the momentum constraints. Also Eqs.~(\ref{sigmaeq}) and (\ref{mueq}) are automatically solved under these conditions. The system of constraints $\mathcal{H} = \mathcal{C} = 0$ given in (\ref{c1}) and (\ref{c2}), for $\alpha \neq 2$, reduces to
\begin{eqnarray}
 \nabla^2 N &=& 0 \,,
\label{cstatic1}
\\
 R + \alpha a_i a^i &=& 0 \,. 
\label{cstatic2}
\end{eqnarray}
The equation of motion (\ref{dotpi}), after inserting (\ref{cstatic1}) and (\ref{cstatic2}), yields
\begin{equation}
 R^{ij} - N^{-1} \nabla^i \nabla^j N + \alpha a^i a^j = 0 \,.
\label{eomstatic}
\end{equation}
The Eqs.~(\ref{cstatic1} - \ref{eomstatic}) are equal to the Lagrangian equations of motion (\ref{hamiltonianfin} - \ref{eomstaticlag}). For $\alpha = 2$, Eq.~(\ref{c1}) gives no information and (\ref{c2}) is the trace of Eq.~(\ref{dotpi}), which in turn matches with the Lagrangian equation of motion (\ref{eomalpha2}).

We conclude this appendix by briefly commenting on how the solution arises in the Hamiltonian formulation of the $\lambda \neq 1/3$ case. The main difference the $\lambda \neq 1/3$ theory has with respect to the $\lambda = 1/3$ one is the absence of the $\pi = 0$ and $\mathcal{C} = 0$ constraints. Consequently, the preservation in time of the second-class constraints ($\phi = \mathcal{H} = 0$) leads to only one equation for $\sigma$. Since static configurations with vanishing shift function have again $\pi^{ij}= 0$, the condition $\pi = 0$ holds anyway and the Hamiltonian constraints of both cases become identical since they differ in general by a term proportional to $\pi^2$. The equation for $\sigma$, which can be found in Ref.~\cite{Bellorin:2011ff}, is totally solved by $\sigma = 0$, which is a consequence of staticity and $N_i = 0$. When all these conditions are imposed on the time evolution of $\pi^{ij}$, the resulting equation is exactly equal to Eq.~(\ref{dotpi}) evaluated on $\pi^{ij} = 0$. Moreover, the trace of this equation is just the $\mathcal{C}$ constraint (\ref{cconstraintquadratic}). Therefore, the constraints/equations of motion of the Hamiltonian formulation of the $\lambda \neq 1/3$ case evaluated on static configurations with $N_i = 0$ lead to the system of Eqs.~(\ref{hamiltonianconstrainquadratic}), (\ref{cconstraintquadratic}) and (\ref{dotpi}) of the $\lambda = 1/3$ case, which is the system of equations we solved in the main body of the paper.

\section{Equivalences between solutions}
\label{apendix:equivalences}
\begin{enumerate}
\item {\bf Case $\alpha < 2$: solutions in $\chi \in (-\infty,0)$}\\
In $\chi\in (-\infty,0)$ one arrives at the same expressions (\ref{coordinatetransf}) and (\ref{nffinalalphal2}) but with negative integration constants,
\begin{equation}
 r = \frac{\tilde{k} e^{s\chi/\beta}}{\sinh{\chi}} \,,
\hspace{2em}
 N = \frac{\beta \tilde{C}}{\tilde{k}} e^{-s\chi/\beta} \,,
\hspace{2em}
 f = (\beta^{-1} \sinh{\chi} - s \cosh{\chi} )^2 \,,
\end{equation}
where $\tilde{C},\tilde{k} < 0$. After the coordinate transformation $\chi' = - \chi$ and the identification of integration constants $\tilde{C} =-C$ and $\tilde{k} = - k$, one gets the two solutions of (\ref{coordinatetransf}) and (\ref{nffinalalphal2}) in $\chi' \in (0,+\infty)$.

\item {\bf Case $\alpha > 2$: Equivalence of the $s=+1$ and $s=-1$ solutions}\\
The $s=-1$ solution is
\begin{equation}
 r = \frac{\tilde{k} e^{-\varphi/\beta}}{\sin{\varphi}} \,,
\hspace{2em}
 N = \frac{\beta \tilde{C}}{\tilde{k}} e^{\varphi/\beta} \,,
\hspace{2em}
 f = (\beta^{-1} \sin{\varphi} + \cos{\varphi} )^2 \,,
\end{equation}
with $\varphi\in (0,\pi)$ and $\tilde{C},\tilde{k} > 0$. If one makes the coordinate transformation $\varphi' = - \varphi + \pi$ and the identification $\tilde{C} = C$, $\tilde{k} = e^{\pi/\beta} k$, one then arrives at the $s=+1$ solution given in (\ref{coordinatetrans2}) and (\ref{fngreaterthan2}).

\item {\bf Case $\alpha > 2$: Solution in $\varphi \in (\pi,2\pi)$}\\
For the $s=+1$ solution in $\varphi \in (\pi,2\pi)$ one arrives at the same expressions (\ref{coordinatetrans2}) and (\ref{fngreaterthan2}) but with negative integration constants,
\begin{equation}
 r = \frac{\tilde{k} e^{\varphi/\beta}}{\sin{\varphi}} \,,
\hspace{2em}
 N = \frac{\beta \tilde{C}}{\tilde{k}} e^{-\varphi/\beta} \,,
\hspace{2em}
 f = (\beta^{-1} \sin{\varphi} - \cos{\varphi} )^2 \,,
\end{equation}
where $\tilde{C},\tilde{k} < 0$. The appropiated transformation is $\varphi' = \varphi - \pi$ and the identification of constants is $\tilde{C} = - C$, $\tilde{k} = - e^{-\pi/\beta} k$. The proof that the $s=-1$ and $s=+1$ solutions in $\varphi \in (\pi,2\pi)$ are equivalent is done in parallel to equivalence 2.

\item {\bf Case $\alpha > 2$: the other four solutions}\\
The other four solutions Eq.~(\ref{bpositive}) has are
\begin{equation}
 \frac{ \beta C' }{r N} = s_1 \cos{\varphi} \,,
\hspace{2em}
 \sqrt{f} + \frac{ C'}{rN} = s_2 \sin{\varphi} \,.
\end{equation}
Among them we first take the $s=+1$ solution in the range $\varphi \in (-\pi/2 , +\pi/2)$. It is
\begin{equation}
 r = \frac{\tilde{k} e^{-\varphi/\beta}}{\cos{\varphi}} \,,
\hspace{2em}
 N = \frac{\beta \tilde{C}}{\tilde{k}} e^{\varphi/\beta} \,,
\hspace{2em}
 f = (\beta^{-1} \cos{\varphi} - \sin{\varphi} )^2 \,,
\label{othersolution}
\end{equation}
with $\tilde{C},\tilde{k} > 0$. The coordinate transformation is $\varphi' = -\varphi + \pi/2$ and the identification is $\tilde{C} = C$, $\tilde{k} = e^{\pi/2\beta} k$. This leads exactly to the solution (\ref{coordinatetrans2}) and (\ref{fngreaterthan2}). By making similar analysis to the previous equivalences one can show that the $s=-1$ solution  in the range $\varphi\in (-\pi/2,+\pi/2)$ and the $s=+1$ and $s=-1$ solutions in $(\pi/2,3\pi/2)$ are all equivalent to (\ref{othersolution}).

\end{enumerate}


\section{Expanding in $\alpha$ the exact solution}
\label{appendix:expanding}
Here we show how the perturbative solution can be obtained from the exact solution when $\alpha \sim 0$. The main point is that the coordinate transformation (\ref{coordinatetransf}) can be explicitly inverted at linear order in $\alpha$. For simplicity, we consider only the positive mass solution ($s=+1$). At the end we shall indicate how to recover the expansion of the negative mass solution ($s=-1$).

We start by rewriting the coordinate transformation (\ref{coordinatetransf}),
\begin{equation}
 r = \frac{\beta C e^{\chi / \beta}}{\sinh{\chi}} \,,
\hspace{2em}
 \chi > 0 \,,
\hspace{2em}
 \beta = \sqrt{ 1 - \alpha / 2 } 
\label{app:coordinatetransf}
\end{equation}
and $C$ is the unique nonfixed integration constant the solution has. We recall that to arrive at (\ref{app:coordinatetransf}) it is assumed $C>0$. When expanded up to linear order in $\alpha$, this transformation becomes
\begin{equation}
 {\displaystyle r =
 2 C \frac{1 + \alpha ( \chi - 1 ) / 4 }{1 - e^{-2\chi}} } \,.
\label{coordinatetranspertur}
\end{equation}
We may further refine this expansion if we assume that $\chi$, as a function of $r$, is of linear order in $\alpha$; that is, we use the ansatz
\begin{equation}
 \chi(r) = \chi^{(0)}(r) + \alpha\: \chi^{(1)}(r) \,.
\label{app:chiexpanded} 
\end{equation}
We first put this expansion back into Eq.~(\ref{coordinatetranspertur}) and evaluate the resulting equation at zeroth order in $\alpha$. This fixes $\chi^{(0)}(r)$,
\begin{equation}
 e^{-2 \chi^{(0)}} = 1 - \frac{2C}{r} \,.
\end{equation}
The second step is to put $\chi^{(0)}(r)$ and the expansion (\ref{app:chiexpanded}) back into Eq.~(\ref{coordinatetranspertur}) and expand up to linear order in $\alpha$. This gives $\chi^{(1)}(r)$ as a combination of $\chi^{(0)}(r)$ and $r$. The whole, linear order function $\chi(r)$ is
\begin{equation}
 \chi =
 - \frac{1}{2} \ln \left( 1 - \frac{2 C}{r} \right)
 - \frac{\alpha C}{8 r} \left( 1 - \frac{2 C}{r} \right)^{-1}
    \left[ 2 + \ln \left( 1 - \frac{2 C}{r} \right) \right] \,.
\end{equation}
The final step is to insert this coordinate transformation into the exact solution, which is given in terms of functions of $\chi$ as
\begin{equation}
 N = e^{ -\chi / \beta} \,,
\hspace{2em}
 f = ( \cosh\chi - \beta^{-1} \sinh\chi )^2\,,
\end{equation}
and expand the functions $N$ and $f$ up to linear order in $\alpha$. We obtain the linear-order solution in terms of the spherical coordinate $r$:
\begin{equation}
\begin{array}{rcl}
 N &=& {\displaystyle
 \left( 1 - \frac{2 C}{r} \right)^{1/2}
 + \frac{\alpha}{8} \left( 1 - \frac{2 C}{r} \right)^{-1/2}
   \left[ \frac{2C}{r}
    + \left( 1 - \frac{C}{r} \right) \ln \left( 1 - \frac{2 C}{r} \right)
     \right] } \,,
\\[2ex]
 f &=& {\displaystyle 
 1 - \frac{2C}{r} 
 + \frac{\alpha C}{4r} \ln\left( 1 - \frac{2 C}{r} \right) 
} \,.
\end{array}
\label{finalexpansion}
\end{equation}
This coincides with the perturbative solution (\ref{perturbativesolution})  if we set the value $B = 0$ and identify the free integration constants of both versions according to $A= 2C$. This implies that we take $A>0$ in (\ref{perturbativesolution}). The expansion of the negative-mass solution can be obtained from (\ref{finalexpansion}) by substituting $C \rightarrow -C$ everywhere in these expressions. This is equivalent to take $A = -2C$, $B = 0$ in (\ref{perturbativesolution}), with $C>0$. 


\end{document}